\UseRawInputEncoding

\documentclass[journal]{IEEEtran}  

\usepackage{bm}
\usepackage{amsmath,amsfonts,amssymb}
\usepackage{algorithmic}
\usepackage{array}
\usepackage[caption=false,font=footnotesize,labelfont=rm,textfont=rm]{subfig}
\usepackage{textcomp}
\usepackage{stfloats}
\usepackage{url}
\usepackage{verbatim}
\usepackage{graphicx}

\usepackage{algorithm}
\usepackage{algorithmic}

\usepackage{color}
\usepackage{amssymb}
\usepackage{booktabs}
\usepackage[table]{xcolor}
\usepackage{graphicx}
\usepackage{multirow}
\usepackage{color} 
\usepackage{cite}

\newcommand{\figref}[1]{Fig. \ref{#1}}

\newcommand{\tabref}[1]{Table \ref{#1}}
\newcommand{\subfigref}[2]{Fig. \ref{#1}\subref{#2}}

\newcommand{\ra}[1]{\renewcommand{\arraystretch}{#1}}

\def\BibTeX{{\rm B\kern-.05em{\sc i\kern-.025em b}\kern-.08em
    T\kern-.1667em\lower.7ex\hbox{E}\kern-.125emX}}
\usepackage{balance}
\usepackage{hyperref}
\hypersetup{hidelinks} 
\hypersetup{
	colorlinks=true,
	linkcolor=blue
}
\begin{document}

%
\title{Near-Field Channel Estimation for XL-MIMO: A Deep Generative Model Guided by Side Information}
\author{
	Zhenzhou~Jin,~\IEEEmembership{Graduate Student Member,~IEEE,}
	Li~You,~\IEEEmembership{Senior Member,~IEEE,}\\
	Derrick Wing Kwan~Ng,~\IEEEmembership{Fellow,~IEEE,}
	Xiang-Gen~Xia,~\IEEEmembership{Fellow,~IEEE,}
	and~Xiqi~Gao,~\IEEEmembership{Fellow,~IEEE}
\thanks{Part of this work has been accepted for presentation at the IEEE GLOBECOM 2024 \cite{jin}.
	
		
		Zhenzhou Jin, Li You, and Xiqi Gao are with the National Mobile Communications Research Laboratory, Southeast University, Nanjing 210096, China, and also with the Purple Mountain Laboratories, Nanjing 211100, China (e-mail: zzjin@seu.edu.cn; lyou@seu.edu.cn; xqgao@seu.edu.cn).
		
		Derrick Wing Kwan~Ng is with the School of Electrical Engineering and Telecommunications, University of New South Wales, Sydney, 2052, Australia (e-mail: w.k.ng@unsw.edu.au).
	
		Xiang-Gen Xia is with the Department of Electrical and Computer Engineering, University of Delaware, Newark, DE 19716, USA (e-mail: xxia@ee.udel.edu).
	}



			
	}

\maketitle

\begin{abstract}
This paper investigates the near-field (NF) channel estimation (CE) for extremely large-scale multiple-input multiple-output (XL-MIMO) systems. Considering the pronounced NF effects in XL-MIMO communications, we first establish a joint angle-distance (AD) domain-based spherical-wavefront physical channel model that captures the inherent sparsity of XL-MIMO channels. Leveraging the channel's sparsity in the joint AD domain, the CE is approached as a task of reconstructing sparse signals. Anchored in this framework, we first propose a compressed sensing algorithm to acquire a preliminary channel estimate. Harnessing the powerful implicit prior learning capability of generative artificial intelligence (GenAI), we further propose a GenAI-based approach to refine the estimated channel. Specifically, we introduce the preliminary estimated channel as side information, and derive the evidence lower bound (ELBO) of the log-marginal distribution of the target NF channel conditioned on the preliminary estimated channel, which serves as the optimization objective for the proposed generative diffusion model (GDM). Additionally, we introduce a more generalized version of the GDM, the non-Markovian GDM (NM-GDM), to accelerate the sampling process, achieving an approximately tenfold enhancement in sampling efficiency. Experimental results indicate that the proposed approach is capable of offering substantial performance gain in CE compared to existing benchmark schemes within NF XL-MIMO systems. Furthermore, our approach exhibits enhanced generalization capabilities in both the NF or far-field (FF) regions. 
\end{abstract}

\section{Introduction}\label{sec:net_intro}
\IEEEPARstart{A}{s} we venture into the sixth-generation (6G) era, two groundbreaking wireless innovations are identified as key enablers for achieving high data rates. The first is extremely large-scale MIMO (XL-MIMO), characterized by the deployment of enormous antenna arrays at the base station (BS), which can significantly enhance spectral efficiency \cite{de2020non,10286475,10403776,an2024near}. The second one is related to the increase in carrier frequencies, to the millimeter wave (mmWave) or terahertz (THz) bands \cite{petrov2020ieee, zhang2023near, JIN_CKM, yang2024near}, which provides a substantial amount of unlicensed available bandwidth. This advancement not only supports the higher data rates regained by 6G networks but also facilitates the implementation of XL-MIMO by allowing for smaller antenna array aperture sizes. Hence, XL-MIMO is increasingly acknowledged as a disruptive technique that will drive the development of upcoming 6G networks.

The efficiency of numerous techniques in XL-MIMO systems, including precoding and beamforming, hinges critically on accurate channel estimation (CE) \cite{10419346,10659325,10440321,jin2024gdm4mmimo,wang2024tutorial}. However, uplink CE encounters various challenges for XL-MIMO orthogonal frequency division multiplexing (OFDM) systems. Considering the hardware costs, practical BSs usually adopt a hybrid digital-analog architecture \cite{9693928,yang2023channel}, where the number of radio frequency (RF) chains is substantially fewer than that of antennas. Consequently, the BS may not be capable of simultaneously observing the received pilot of each antenna, resulting in increased communication overhead. More pressing is the issue that the dramatic increase in antenna count within XL-MIMO systems will lead to unaffordable pilot signaling overhead. Therefore, achieving accurate CE while concurrently minimizing pilot signaling overhead is paramount.

In current fifth-generation (5G) massive MIMO systems, the majority of enabling techniques rely on the far-field (FF) assumption that electronic wave radiation adheres to the planar-wavefront model \cite{9617121,gong2024near}. These techniques leverage the channel's sparsity representation in the angular domain to estimate the high-dimensional channel \cite{dai2018fdd,fan2018angle,8611231,9305993}. Nevertheless, this assumption might not hold in XL-MIMO systems. In particular, the Fraunhofer distance is given by ${d_{\rm{F}}} = {{2{D^2}} \mathord{\left/
		{\vphantom {{2{D^2}} \lambda }} \right.
		\kern-\nulldelimiterspace} \lambda }$, where $D$ is the aperture of the antenna array, and $\lambda$ is the wavelength of the signal \cite{zhang2023near}. Therefore, as the number of antennas rises, the Fraunhofer distance, which serves as the boundary between the FF and near-field (NF) regions, also increases \cite{7942128}. Consequently, more user terminals will fall within the NF region, where electromagnetic waves exhibit as spherical waves rather than adhere to the characteristics of FF propagation. Additionally, to more accurately characterize the near-far field boundary, the authors of \cite{ZengShuhao} considered the cross-polarization discrimination (XPD) variances for dual-polarized XL-MIMO systems and further introduced the non-uniform XPD distance to complement existing near-far field boundary. Under such circumstances, applying traditional FF assumption-based CE methods, e.g., \cite{dai2018fdd,fan2018angle}, to emerging NF scenarios will experience performance degradation due to the pronounced spherical-wavefront property. Hence, in the context of XL-MIMO communications, adopting a spherical-wavefront model becomes more pertinent, especially since the time difference of arrivals between two antennas is influenced by both the distances and the angles of arrivals \cite{9693928,zhang2023near}.

Due to the unique characteristics of the spherical-wavefront channel, NF XL-MIMO necessitates exploiting sparsity in both angle and distance dimensions for efficient CE \cite{9693928}. Considering the sparsity of the channel, several works have studied the XL-MIMO CE problem. For instance, the authors of \cite{9693928} proposed a new sparse representation in the polar domain, effectively capturing both distance and angular information, and then adopted the simultaneous orthogonal matching pursuit (SOMP) algorithm to acquire the channel parameters. Additionally, two typical XL-MIMO scenarios, hybrid-field and combined line-of-sight (LoS) and non-line-of-sight (NLoS) NF, were investigated in \cite{9598863} and \cite{10078317}, respectively, where the OMP algorithm was adopted in both scenarios. With the further development of holographic MIMO technology, the authors of \cite{10620366} thoroughly investigated the power diffusion effect in the sparse-signal-recovery-based hybrid-field CE and accordingly proposed a power-diffusion-aware hybrid-field CE method (PD-OMP) for holographic MIMO communications. Note that most of the aforementioned methodologies are based on compressed sensing (CS), which assumes that the angles and distances lie on discrete points in the polar domain (i.e., on-grid angles and distance). In contrast, the actual angles and distances are continuously distributed, i.e., off-grid angles and distances. Consequently, the estimation accuracy of such methods may be limited \cite{alkhateeb2014channel,9693928,zhang2023near}.
To this end, various off-grid methods have been proposed \cite{ma2020high,hu2018super,iimori2022joint}. However, the off-grid algorithms require a significant trade-off between accuracy and complexity, highlighting an ongoing challenge in developing efficient CE techniques in XL-MIMO systems.

In response to these challenges, deep learning (DL)-based CE has recently been proposed to excavate the intrinsic characteristics of the channel in NF XL-MIMO systems \cite{9491074,he2018deep,lei2023channel}. For example, \cite{he2018deep} introduced the learned denoising-based approximate message passing (L-DAMP) algorithm, which incorporates a convolutional neural network into the iterative recovery framework of AMP. In \cite{lei2023channel}, a CE method based on a multi-layer residual network was proposed to capture the channel's sparsity and fuse multi-scale features. However, most feedforward neural network-based regression methods for CE often perform poorly because the conditional distribution of the estimated channel outputs for a given pilot signal or partial measurement channel input typically does not follow a simple parametric distribution.

With the advent of the generative AI (GenAI) era, deep generative models such as variational autoencoders (VAEs) \cite{kingma2013auto}, generative adversarial networks (GANs) \cite{goodfellow2020generative}, energy-based models (EBMs) \cite{10419041}, and normalizing flows (NFs) \cite{kobyzev2020normalizing} have demonstrated robust latent representations by learning data-driven implicit priors. Notably, generative diffusion models (GDMs) \cite{Ho11} are recognized as one of the most paramount class of the generative models, which do not require training additional discriminators akin to GANs, aligning posterior distributions like VAEs, or enforcing network constraints as NFs. Besides, GDMs alleviate the computational burdens associated with Markov chain Monte Carlo methods during training in EBMs \cite{10419041}. Therefore, GDMs have garnered growing attention for their robust capabilities to capture intricate relationships and distributions, facilitating the synthesis and reconstruction of high-dimensional target data \cite{Ho11, Rombach_2022_CVPR}. GDMs represent a class of probabilistic generative models, where the reverse denoising process can be interpreted as solving an inverse problem. With powerful implicit prior learning capabilities, GDMs excel in modeling complicated data distributions and providing robust prior information, thereby mitigating the performance degradation caused by mismatches between handcrafted priors and real-world scenarios. Their exceptional performance in multi-modal learning tasks, such as SORA and Stable Diffusion \cite{Rombach_2022_CVPR}, highlights their versatile application potential and efficacy. Specifically, a GDM trains a denoising neural network by minimizing the Kullback-Leibler divergence between the predicted and target distributions. This approach enables the model to learn to transform a standard normal distribution into an empirical data distribution through a series of refinement steps. 

To effectively mitigate the high-dimensional inverse problem in NF CE for XL-MIMO systems, a pre-trained GDM can be leveraged as a data-driven prior by providing the score function of the underlying channel distribution, without assuming specific structures of the distribution. The score function of a distribution, defined as the gradient of its log-probability, has been demonstrated to be equivalent to the noise learned by GDM \cite{luo2022understanding}. If the implicit prior information of the target channel distribution, such as the score function, can be learned, one can transition to the target channel distribution from a standard normal distribution through iterative refinement steps, akin to Langevin dynamics \cite{song2019generative}. Given the challenges posed by our considered scenario, GDM, widely recognized for its strong implicit prior learning capabilities, naturally emerges as a promising candidate for XL-MIMO NF CE. However, the conventional GDM, through a diffusion-inverse diffusion process, continuously learns and fits the distribution of the target data during denoising, gradually generating a channel sample that approximately follows the target distribution. In other words, the conventional GDM learns only the distribution $p(\mathbf{h})$ of the target channel, rather than directly solving for $p(\mathbf{h}|\mathbf{y})$ (the posterior estimation), where $\mathbf{y}$ represents the observed or side information. Since the conventional GDM does not incorporate $\mathbf{y}$ to refine the generated channel ${{\mathbf{\hat h}}}$, the resulting ${{\mathbf{\hat h}}}$ is simply a sample drawn from the learned target distribution, and may not accurately represent the true channel ${{\mathbf{ h}}}$ in the current environment. Therefore, the conventional GDM functions merely as a channel generator that may not generate an accurate-enough channel, rather than a precise channel estimator.


Based on the above analysis, we propose a GDM \textit{conditioned} on side information for NF CE in an uplink XL-MIMO OFDM system. Specifically, the proposed approach is a two-stage CE framework, and the main contributions of this study are summarized as follows:

\begin{itemize}
	
	\item Considering the pronounced NF effects, we transform the channel representation into the joint angle-distance (AD) domain and leverage its sparsity to introduce the SOMP algorithm for acquiring an initial CE. To facilitate subsequent processing by the proposed neural networks, the real and imaginary parts of the initial CE are separated and concatenated into a multidimensional tensor, resembling the structure of a color image.
	
	\item To further refine the initial CE, we tailor a GDM conditioned on side information. Specifically, we leverage variational inference techniques to derive the evidence lower bound (ELBO) of the log-marginal distribution of the NF channel conditioned on the initial CE, serving as a proxy objective. Employing this proxy objective, the trained GDM takes the initial CE and a noise image as inputs to perform a series of iterative refinement steps, akin to Langevin dynamics.
	
	\item To accelerate the generation process of the GDM conditioned on the initial CE, we reveal the inherent limitations and introduce a non-Markovian chain-like decomposition structure, referred to as NM-GDM, to enhance sampling efficiency. Meanwhile, we further analyze that both NM-GDM and GDM conditioned on the initial CE share the same objective function, meaning that the NM-GDM is a more generalized form of the GDM. 
	
	\item Experimental results show that the proposed approach achieves competitive CE performance and robust generalization capability compared to the baseline methods, even under conditions with limited communication overhead. Additionally, our approach exhibits impressive CE performance in low-SNR conditions by leveraging learned implicit prior information to mitigate the impact of noise.

\end{itemize}

The rest of this paper is structured as follows. Section II introduces the XL-MIMO NF system model. The GenAI-aided CE methodology is proposed in Section III. In Section IV, the network architecture of the proposed NM-GDM is introduced. Experimental results are presented in Section V, with the conclusions provided in Section VI. 

\emph{Notations}: ${\jmath=\sqrt{-1}}$ represents the imaginary unit; ${\left(  \cdot  \right)^T}$ and ${\left(  \cdot  \right)^H}$ represent the transpose and Hermitian transpose operators, respectively. ${\cal O}\left( \cdot \right)$ represents the remainder terms in the Taylor expansion formula. $\mathbb{C}^{N \times M}$ ($\mathbb{R}^{N \times M}$) represents the space of $N \times M$ dimensional complex (real) matrix. ${\mathbf{I}}_N$ represents the identity matrix of dimension $N \times N$. ${{{\mathbf{C}}}\left( {i,j} \right)}$ indicates the element at position $(i,j)$ in matrix ${{\mathbf{C}}}$. $\mathbb{E}_q\left( \cdot \right)$ represents the expectation operation w.r.t. the distribution $q(x)$. $\propto$, $\rm{tr}\left( \cdot \right)$, and $\rm{det}\left( \cdot \right)$ represent the proportional, trace, and determinant operators, respectively. ${\left\|  \cdot  \right\|_{{{2}}}}$ represents the $l_2$-norm. ${\rm{Bdiag}}\left(\mathbf{D}_1,\mathbf{D}_2,...\right)$ represents a block diagonal matrix where the matrix blocks ${\mathbf{D}}_i$ are positioned along the main diagonal. $\odot$ denotes element-wise multiplication. $\Re \mathfrak{e}\left(  \cdot  \right)$ and $\Im \mathfrak{m}\left(  \cdot  \right)$ represent the real and imaginary components of a complex number (or matrix), respectively. $\mathbf{Y}[i,j]$, $\mathbf{Y}[i,:]$, and $\mathbf{Y}[:,j]$ represent the element at position $(i, j)$, all the elements in the $i$-th row, and all the elements in the $j$-th column of matrix $\mathbf{Y}$, respectively. $\mathcal{CN}\left(x;\mu,v \right)$ represents the complex Gaussian distribution characterized by a mean of $\mu$ and a variance of $v$. ${\left[ {{{{\mathbf{G}}}}} \right]_{:,:,1}}$ denotes all the elements located at the first index position along the third dimension of tensor $\mathbf{G}$. $D_{\rm{KL}}(\cdot)$ is the Kullback-Leibler (KL) divergence.
\section{System Model}
\begin{figure}[!t]
	\centering
	\includegraphics[scale=0.41]{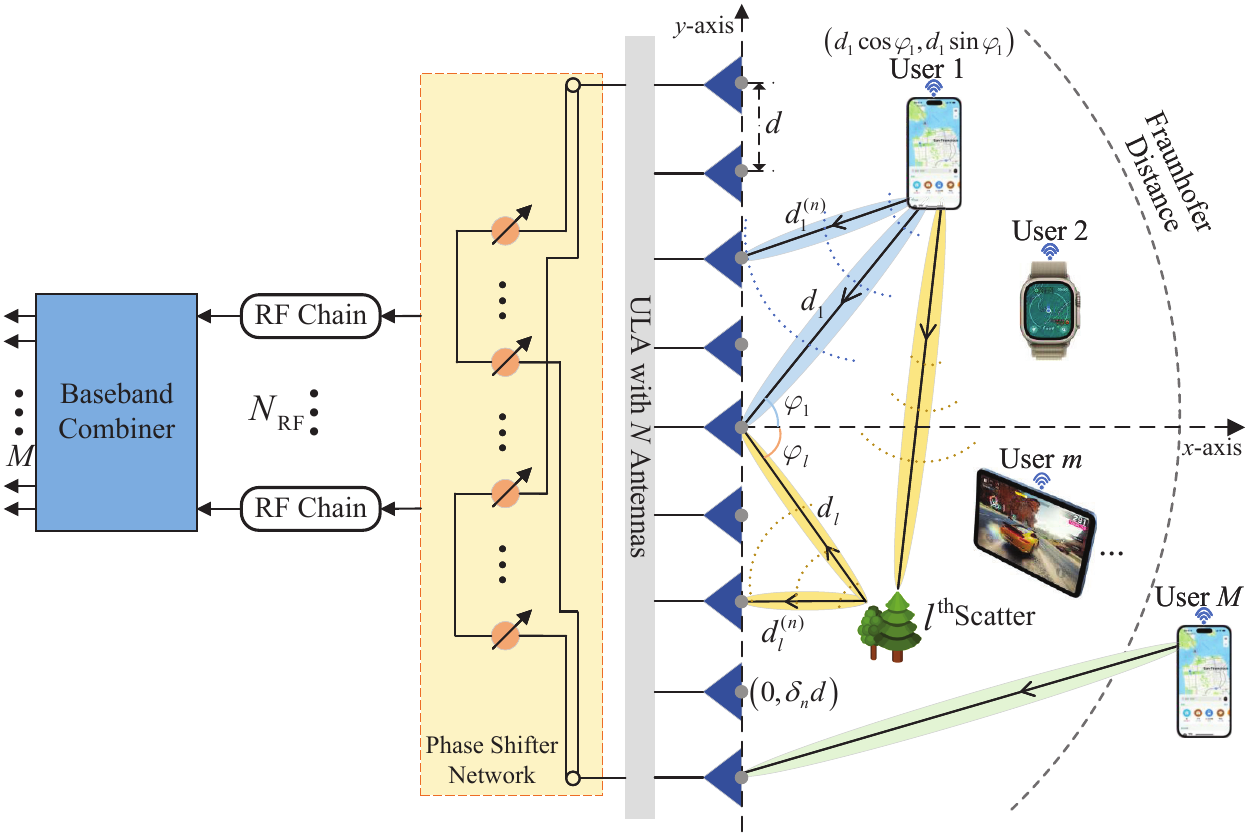}
	\captionsetup{font=footnotesize}
	\caption{Diagram of the NF XL-MIMO system.}
	\label{fig:antenna}
\end{figure}

In this section, we start with the spherical-wavefront channel model. Then, a joint AD domain-based channel model is introduced. Ultimately, the received signal model is depicted with the joint AD domain channel model. Specifically, consider a multi-user uplink XL-MIMO OFDM system in which the BS is outfitted with a uniform linear array (ULA) of $N$ antennas, spaced at half-wavelength intervals, and the $M$ users are single-antenna devices. To reduce power consumption and hardware costs, the BS adopts a hybrid digital-analog architecture with a restricted number, $N_{\rm{RF}}$ ($< N$), RF chains \cite{8306126,alkhateeb2014channel}. As shown in \figref{fig:antenna}, the ULA is assumed to be oriented along the $y$-axis and the $x$-$y$ coordinates of each antenna are denoted as $\left( 0,{{{\delta _n}}d} \right)$, where $d={\lambda _c}/2$ denotes the antenna spacing and ${\lambda _c}$ represents the carrier wavelength, ${\delta _n} = ({{2n - N + 1}})/{2}$, $n=0,1,...,N-1$. We focus on the CE for an uplink XL-MIMO OFDM system, and the $M$ users are assumed to transmit mutually orthogonal pilot sequences to the BS. For the sake of generality, we analyze an arbitrary user located at $\left( {{d_1}\cos {\varphi _1},{d_1}\sin {\varphi _1}} \right)$, where ${d_1}$ and ${\varphi _1}$ denote the distance and angle from the user to the center of the BS antenna array, respectively. Note that $l=1$ and $l\geq 2$ represent the LoS and NLoS path components, respectively.
\subsection{Spherical Wavefront Channel Model}\label{subsec:chmod}
Electromagnetic energy radiates from the antenna and propagates through free space as a series of expanding spherical wavefronts \cite{zhang2023near}. Hence, the channel ${\mathbf{h}_{{k}}}$ from a particular user to the BS at the $k$-th subcarrier is generally modeled as \cite{1137900}
\begin{align}\label{eq:hk}
	{{\mathbf{h}}_{k}} = {\sqrt {\frac{{{N}}}{L}} }\sum\limits_{l = 1}^L {g_l}{e^{ - {\jmath}n_k{d_l}}}{\mathbf{\dot a_B}}\left( {{\varphi _l},{d_l}} \right),
\end{align}
where ${n_k} = {{2\pi {f_k}} \mathord{\left/
		{\vphantom {{2\pi {f_k}} c}} \right.
		\kern-\nulldelimiterspace} c}$, $c$ and $f_k$ are the light speed and carrier frequency, respectively. ${g_l}$ and ${\varphi _l}$ are the complex channel gain and angle of arrival on the $l$-th path, respectively. $L$ is the number of propagation paths. Generally, ${{{\bf{\dot a}}}_{\bf{B}}}\left(  \cdot  \right)$ is considered a frequency-independent steering vector \cite{zhang2023near} based on the NF spherical wavefront assumption, expressed as \cite{9693928}
\begin{align}\label{eq:ab}
   {{{\bf{\dot a}}}_{\bf{B}}}\!\left( {{\varphi _l},{d_l}} \right) \!=\! \frac{1}{{\sqrt {{N}} }}{\left[ {{e^{ - \jmath{\textstyle{{2\pi } \over \lambda_c }}\!\left(\! {d_l^{(0)} \!- {d_l}} \!\right)}},...,{e^{ - \jmath{\textstyle{{2\pi } \over \lambda_c }}\!\left(\! {d_l^{(N \!-\! 1)} \!- {d_l}} \!\right)}}} \right]^T}\!,
\end{align}
where ${{d_l}}$ and ${d_l^{(n)}}$ represent the distance from the center of the BS antenna array and the $n$-th BS antenna element to the user or scatterer on the $l$-th path, respectively. $d_l^{(n)}$ is given by $d_l^{(n)} \!\!\!=\!\!\! \sqrt {{d_l}^2 + \delta _{n}^2{d^2} - 2{d_l}{\delta _{n}}d\sin {\varphi _l}}$, where ${\varphi _l} \in [{{ - \pi } \mathord{\left/
		{\vphantom {{ - \pi } 2}} \right.
		\kern-\nulldelimiterspace} 2},{\pi  \mathord{\left/
		{\vphantom {\pi  2}} \right.
		\kern-\nulldelimiterspace} 2}]$ is the actual physical angle. Based on the spherical wavefront propagation assumption, $( d_l^{(n)} - {d_l}) $ in \eqref{eq:ab} can be approximated as $d_l^{(n)} - {d_l} \approx {{\delta _n^2{d^2}{{\cos }^2}{\varphi _l}} \mathord{\left/
		{\vphantom {{\delta _n^2{d^2}{{\cos }^2}{\varphi _l}} {2{d_l}}}} \right.
		\kern-\nulldelimiterspace} {2{d_l}}} - {\delta _n}d\sin {\varphi _l}$, which is derived by $\sqrt { {1 + z}}  \approx 1 + {z \mathord{\left/
		{\vphantom {z 2}} \right.
		\kern-\nulldelimiterspace} 2} - {{{z^2}} \mathord{\left/
		{\vphantom {{{z^2}} 8}} \right.
		\kern-\nulldelimiterspace} 8} + {\cal O}\left( {{z^2}} \right)$ when $z$ is close to 0. It is sufficiently precise in the NF region when the communication distance is exceeds the Fresnel boundary ${d_{{\rm{FR}}}} = {D\sqrt{D/{{\lambda}_{c}}}}/2$ \cite{9693928}, where $D=Nd$.

\textit{Remark 1:} When the Taylor series expansion only maintains the first-order term, i.e., $\sqrt {{1 + z}}  \approx 1 + z/2$, the difference in distance $(d_l^{(n)}\!-\!d_l)$ becomes proportional to the index $\delta _n$ of the BS antenna, which means that the spherical-wavefront steering vector degenerates into the planar-wavefront one, denoted as $\mathbf{a_B}\left( {\varphi_l} \right) \!=\! \sqrt {1/N} {\left[ {1,{e^{\jmath\pi \sin {\varphi _l}}},...,{e^{\jmath\left( {{N} - 1} \right)\pi \sin {\varphi _l}}}} \right]^T}$. Therefore, the conventional planar-wavefront steering vector can be viewed as a special case of the considered spherical-wavefront counterpart.

Considering the sparsity of the NF spherical-wavefront channel in the joint AD domain \cite{1137900}, the NF channel ${{\bf{h}}_{k}}$ in \eqref{eq:hk} can be converted into its joint AD domain utilizing a new transform matrix $\mathbf{P}$ \cite{9693928}, which is given by
\begin{align}\label{eq:chmd}
  \begin{split}
	\mathbf{P} &= \left[ {{{\bf{\dot a}}}_{\bf{B}}}\big( {{\varphi _1},d_1^1} \big),...,{{{\bf{\dot a}}}_{\bf{B}}}\big( {{\varphi _1},d_1^{{S_1}}} \big), ...,\right. \\ 
	&\qquad\left.{{{\bf{\dot a}}}_{\bf{B}}}\big( {{\varphi _{N}},d_{N}^1} \big),...,{{{\bf{\dot a}}}_{\bf{B}}}\big( {{\varphi _{N}},d_{N}^{{S_{N}}}} \big) \right],
  \end{split}
\end{align}	
where each column in $\mathbf{P}$ is the array response vector sampled at the grid coordinates $({\varphi _n}, d_n^{{s_n}})$, and ${s_n} =1,...,{S_n}$. ${S_n}$ is the number of sampled distances at the sampled angle ${{\varphi _n}}$. Therefore, the total count of sampled grids is denoted as $S = \sum\nolimits_{n = 1}^N {{S_n}}$. Utilizing the matrix $\mathbf{P}\in {\mathbb{C}^{N \times S}}$, the channel ${\mathbf{h}_{k}} $ can be represented as
\begin{align}\label{eq:chmd}
{{\bf{h}}_{k}} = \mathbf{P}{\bf{h}}_{k}^{\rm{AD}},
\end{align}
where ${\bf{h}}_{k}^{\rm{AD}}\in {\mathbb{C}^{S \times 1}}$ is the representation of ${\mathbf{h}_{k}}$ in the AD domain and is typically sparse \cite{1137900,9693928}.

\subsection{Signal Model}\label{}
Let ${s_{k,q}}\in \mathbb{C}$ be the pilot symbol transmitted on the $k$-th subcarrier in the $q$-th time slot. Then, for the $k$-th subcarrier, the received pilot ${{\bf{r}}_{k,q}} \in {\mathbb{C}^{N_{\rm{RF}} \times 1}}$ at the BS is 
\begin{align}\label{eq:chmd}
	{\mathbf{r}_{k,q}} = {\mathbf{C}_q}{\mathbf{P}}{\bf{h}}_{k}^{\rm{AD}}{s_{k,q}} + {\mathbf{C}_q}{\mathbf{n}_{k,q}},
\end{align}
where ${\mathbf{n}_{k,q}} \in {\mathbb{C}^{N \times 1}}$ is the additive white Gaussian noise following $\mathcal{CN}\left( {\mathbf{0},{\sigma ^2}{{\mathbf{I}}_N}} \right)$, and ${\sigma ^2}$ denotes the noise power. ${\mathbf{C}_q} \in {\mathbb{C}^{N_{\rm{RF}} \times N}}$ represents the combining matrix with $\left| {{{\mathbf{C}}_q}\left( {i,j} \right)} \right| = 1/{{\sqrt N }}$. When $N_{\rm{RF}}=N$, it is also applicable to the fully digital architecture. Given that the pilot symbol ${s_{k,q}}$ is known at the receiver, for simplification, let $Q$ be the pilot length and assume ${s_{k, q}} =1 $ for $q =1,2,... , Q$ \cite{1137900,zhang2023near}. By gathering the received pilot symbols in $Q$ time slots, we have the collective received pilot sequence ${{\bf{r}}_k} = {\left[ {{\bf{r}}_{_{k,1}}^T,{\bf{r}}_{_{k,2}}^T,...,{\bf{r}}_{_{k,Q}}^T} \right]^T}$ on the $k$-th subcarrier at the BS, which can be represented as 
\begin{align}\label{eq:9}
 {{\bf{r}}_k} = {\mathbf{C}}{{\bf{Ph}}}_{{k}}^{{\rm{AD}}} + {{\bf{n}}_k},
\end{align}  
where ${\mathbf{C}} = {[{\mathbf{C}}_1^T,...,{\mathbf{C}}_Q^T]^T} \in {\mathbb{C}^{Q{N_{{\text{RF}}}} \times N}}$, ${{\mathbf{n}}_k} = {\left[ {{\mathbf{n}}_{k,1}^T{\mathbf{C}}_1^T,{\mathbf{n}}_{k,2}^T{\mathbf{C}}_2^T,...,{\mathbf{n}}_{k,Q}^T{\mathbf{C}}_Q^T} \right]^T}$. 
\section{GenAI-Aided Channel Estimation}
\begin{figure}[!t]
	\centering
	\includegraphics[scale=0.63]{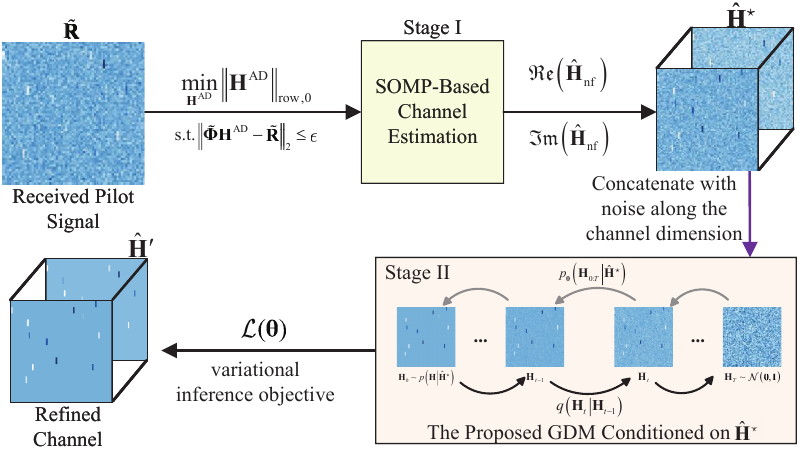}
	\captionsetup{font=footnotesize}
	\caption{Framework of our proposed two-stage NF CE scheme for XL-MIMO. The proposed GDM is conditioned on ${{{{\mathbf{\hat H}}}^ \star }}$, and its implementation involves concatenating the side information ${{{{\mathbf{\hat H}}}^ \star }}$ with pure Gaussian noise ${{\mathbf{H}}_T} \sim \mathcal{N}\left( {\mathbf{0},{\mathbf{I}}} \right)$ along the channel dimension, which then guides a series of iterative refinement steps.}
	\label{fig:liuchengtu}
\end{figure}

In this section, a two-stage CE scheme for NF XL-MIMO systems is proposed, as illustrated in \figref{fig:liuchengtu}. Specifically, in the first stage, considering that the representation of the XL-MIMO NF channel in the joint AD domain exhibits a sparse property, we utilize the SOMP-based approach to acquire a initial CE. Then, we propose a GDM conditioned on the initial CE as side information to enhance CE accuracy. Besides, to accelerate the generation process, we introduce the non-Markovian GDM, conditioned on the initial CE, for efficiently estimating the XL-MIMO channel.
\subsection{Initial CE: CS-Based Near-Field CE}
Considering that the representation of NF channels in the AD domain is sparse \cite{zhang2023near,9693928}, the CE task is capable of reformulating as a task of sparse signal reconstruction. Building upon this CS framework, we further rewrite the received pilot signal $\mathbf{r}_k$ on the $k$-th subcarrier in the general form, i.e.,
\begin{align}\label{eq:21}
	{\mathbf{r}}_k = \boldsymbol{\Phi}{{\mathbf{h}}^{{\rm{AD}}}_{k}} + {\mathbf{n}_k}, \text{ } \forall k,
\end{align}
where $\boldsymbol{\Phi}  = {\mathbf{CP}} \in {\mathbb{C}^{Q{N_{{\text{RF}}}} \times S}}$ represents the sensing matrix. Considering that the CS framework works effectively under the premise that the noise covariance matrix is diagonal, implying no correlation among different noise components \cite{ 8306126}. However, since the noise ${\mathbf{n}}_k$ in the received signal in \eqref{eq:21} is colored, we need to perform a whitening preprocessing operation for effective signal processing in the sequel. Specifically, let ${\boldsymbol{\Sigma} _c} = \mathbb{E}\left( {{{\mathbf{n}}_k}{\mathbf{n}}_k^H} \right) = {\rm{Bdiag}}\left( {{\sigma ^2}{{\mathbf{C}}_1}{\mathbf{C}}_1^H,{\sigma ^2}{{\mathbf{C}}_2}{\mathbf{C}}_2^H,...,{\sigma ^2}{{\mathbf{C}}_Q}{\mathbf{C}}_Q^H} \right)$ be the covariance matrix of the noise ${\mathbf{n}}_k$ in the received signal. According to Cholesky whitening \cite{8306126}, ${\boldsymbol{\Sigma} _c}$ can be decomposed into ${{\mathbf{\Sigma }}_c} = {\sigma ^2}{\boldsymbol{\Xi }}{{\boldsymbol{\Xi }}^H}$, where ${\boldsymbol{\Xi }} \in {\mathbb{C}^{{N_{{\rm{RF}}}}Q \times {N_{{\rm{RF}}}}Q}}$ is a lower triangular matrix and ${{\boldsymbol{\Xi }}^{ - 1}}$ is the pre-whitening matrix. Therefore, the whitened received signal ${{\mathbf{\tilde r}}_k}$ is given by
\begin{align}\label{eq:22}
    {{\mathbf{\tilde r}}_k} = {\mathbf{\tilde \Phi }}{{\mathbf{h}}^{{\rm{AD}}}_{k}}+ {\mathbf{\tilde n}_k},
\end{align}
where ${\mathbf{\tilde \Phi }} = {{\boldsymbol{\Xi}} ^{ - 1}}{\mathbf{\Phi }} = {{\boldsymbol{\Xi}} ^{ - 1}}{\mathbf{CP}}$, ${\mathbf{\tilde n}_k} = {{\mathbf{\Xi }}^{ - 1}}{\mathbf{n}_k}$ and the covariance matrix of ${\mathbf{\tilde n}_k}$ is ${{{\mathbf{\tilde \Sigma }}}_c} = {{\mathbf{\Xi }}^{ - 1}}{\sigma ^2}{\mathbf{\Xi }}{{\mathbf{\Xi }}^H}{{\mathbf{\Xi }}^{ - H}} = {\sigma ^2}{\mathbf{I}_{{N_{\rm{RF}}}Q}}$.

In general, the physical propagation characteristics of the channel within the system bandwidth remain relatively unchanged. Hence, the subchannels corresponding to different subcarriers have similar scatterers in the physical propagation environment \cite{8284057}. Then, the steering vectors ${{{\bf{\dot a}}}_{\bf{B}}}\left( \cdot \right)$ are identical across different subcarriers \cite{9693928}. Consequently, the sparsity support of ${\bf{h}}_{k}^{\rm{AD}}$ are also the same for different subcarriers $\left[ {{f_k}} \right]_{k = 1}^K$, i.e.,
\begin{align}\label{eq:sup}
    {\mathcal{S}} \!=\! {\rm{supp}}\left\{ {{\mathbf{h}}_{1}^{{\rm{AD}}}} \right\} \!=\! {\rm{supp}}\left\{ {{\mathbf{h}}_{2}^{{\rm{AD}}}} \right\} \!=\! ... \!=\! {\rm{supp}}\left\{ {{\mathbf{h}}_{K}^{{{\rm{AD}}}}} \right\}.
\end{align}
Hence, they can be estimated simultaneously to improve the CE accuracy. Then, \eqref{eq:22} can be rearranged as
\begin{align}\label{eq:jieshou}
	{{\mathbf{\tilde R}}} = {\mathbf{\tilde \Phi }}{{\mathbf{H}}^{\rm{AD}}} + {\mathbf{\tilde N}},
\end{align} 
where ${\mathbf{\tilde R}} = \left[ {{{\mathbf{\tilde r}}_1},{{\mathbf{\tilde r}}_2},...,{{\mathbf{\tilde r}}_K}} \right]$, ${{\mathbf{H}}^{{\rm{AD}}}} = [{\mathbf{h}}_{1}^{{\rm{AD}}},{\mathbf{h}}_{2}^{{\rm{AD}}},...,{\mathbf{h}}_{K}^{{\rm{AD}}}]$ and ${\mathbf{\tilde N}} = [{{\mathbf{\tilde n}}_1},{{\mathbf{\tilde n}}_2},...,{{\mathbf{\tilde n}}_K}]$. We aim to acquire the channel ${{\bf{H}}} = \mathbf{P}{\bf{H}}^{\rm{AD}}$ utilizing the received signal matrix ${{\mathbf{\tilde R}}}$ and the sensing matrix ${\mathbf{\tilde \Phi }}$. Combining the sparsity of the channel in the AD domain and its properties in \eqref{eq:sup}, the rows of ${{\mathbf{H}}^{{\rm{AD}}}}$ are sparse. Thus, given the measurement of ${{\mathbf{\tilde R}}}$, the AD domain channel can be obtained by solving the following optimization problem
\begin{subequations}\label{eq:youhua}
	\begin{align}
		&\mathop {\min }\limits_{{\mathbf{H}}^{\rm{AD}}} {\left\| {{{\mathbf{H}}^{{\text{AD}}}}} \right\|_{\rm{row},0}}\label{eq:}\\
		&\ {\mathrm{s.t.}}\quad\!\!\!\!{\left\| {{\mathbf{\tilde \Phi }}{{\mathbf{H}}^{{\text{AD}}}} - {\mathbf{\tilde R}}} \right\|_2} \leq {\epsilon},\text{ and }\eqref{eq:sup},\label{eq:}
	\end{align}
\end{subequations}
where ${\left\|  \cdot  \right\|_{{\rm{row}},{{0}}}}$ denotes the row $l_0$-norm. Note that \eqref{eq:youhua} is a structured sparse signal reconstruction problem, and classical algorithm such as SOMP \cite{8306126} can be adopted to handle it.
\subsection{Enhanced CE: GDM-Assisted CE Refinement}
The SOMP approach assumes that the angles and distances precisely align with sampling points in the joint AD domain, whereas distance and angles are continuous in reality. The inherent limitations may degrade the effectiveness of NF CE \cite{zhang2023near, 9693928}. Therefore, we need to further refine the estimated channel. Specifically, given the AD domain channel based on CS, the channel can be obtained by
\begin{align}\label{eq:error}
   {{{\mathbf{\hat H}}}} = {\mathbf{P}}{{{\mathbf{\hat H}}}^{{\rm{AD}}}} = {{\mathbf{H}}} + \mathbf{e},
\end{align}
where ${{{\mathbf{\hat H}}}^{{\rm{AD}}}}$ is the AD domain channel obtained by the SOMP algorithm, $\mathbf{e}$ denotes the estimation error, and ${{\mathbf{H}}}$ is the ground-truth channel. To further eliminate $\mathbf{e}$, we introduce ${{{\mathbf{\hat H}}}}$ as \textit{side information} to guide the GDM in refining the initial CE. 

\textit{1) Overall Design of GDM Conditioned on Side Information:} GDM aims to learn the implicit prior of the target data distribution, denoted as $q\left( {{{\mathbf{x}}_0}} \right)$, such as ``noise'', which is equivalent to the gradient of the empirical data log-density, and then constructs a tractable distribution ${p_{\boldsymbol{\theta}} }\left( {{{\mathbf{x}}_0}} \right)$ to effectively approximate $q\left( {{{\mathbf{x}}_0}} \right)$. In the second stage of NF CE, we focus on learning a parametric approximation to $p\left( {{\mathbf{H}}}\left| {{{\mathbf{\hat H}}}}\right.\right)$ through a series of guided iterative refinement steps \textit{conditioned} on ${{{\mathbf{\hat H}}}}$. 

Unlike traditional GDMs, we design a \textit{conditional} GDM that incorporates ${{{\mathbf{\hat H}}}}$ as side information for the refinement of NF CE. Specifically, considering the inherent correlation between the real and imaginary parts of the channel, we convert the estimated channel ${{{\mathbf{\hat H}}}}$ into a 2-channel image ${{{{\mathbf{\hat H}}}^ \star }}$, i.e., ${\left[ {{{{\mathbf{\hat H}}}^ \star}} \right]_{:,:,1}} = \Re \mathfrak{e}\left( {{{{\mathbf{\hat H}}}}} \right)$ and ${\left[ {{{{\mathbf{\hat H}}}^ \star }} \right]_{:,:,2}} = \Im \mathfrak{m}\left( {{{{\mathbf{\hat H}}}}} \right)$, respectively. As shown in \figref{fig:liuchengtu}, the proposed GDM can generate the target NF channel ${{\mathbf{H}}}$, hereafter referred to as ${{\mathbf{H}}}_0$ for simplicity, through $T$ refinement time steps. Starting with ${{{\mathbf{H}}}}_T \sim \mathcal{N}\left( {\mathbf{0},\mathbf{I}} \right)$, which is composed of pure Gaussian noise, the proposed GDM iteratively refines this initial input based on the source signal ${{{{\mathbf{\hat H}}}^ \star }}$ and the prior information learned during training, specifically through the conditional distribution ${p_{{\boldsymbol{\theta}}} }\left( {{{{{\mathbf{H}}}}_{t - 1}}\left| {{{{{\mathbf{H}}}}_t},{{{{\mathbf{\hat H}}}^ \star }}} \right.} \right) $. As it progresses through each time step $t$, it produces a series of intermediate channels defined as $\left\lbrace {{{\mathbf{H}}}}_{T-1}, {{{\mathbf{H}}}}_{T-2},...,{{{\mathbf{H}}}}_{0} \right\rbrace $, ultimately resulting in the target NF channel ${{{\mathbf{H}}}}_0 \sim p\left({{{\mathbf{H}}}} \left| { {{{{\mathbf{\hat H}}}^ \star }}} \right. \right) $. In principle, the distribution of the intermediate channels in the iterative refinement process is determined by the forward diffusion process, which progressively introduces noise into the output channel via a fixed Markov chain, represented as ${q\left( {{{\mathbf{H}}_t}\left| {{{\mathbf{H}}_{t - 1}}} \right.} \right)}$. 

Our model seeks to reverse the Gaussian diffusion process by iteratively recovering the signal from noise through a reverse Markov chain \textit{conditioned} on the side information ${{{{\mathbf{\hat H}}}^ \star }}$. To achieve this, we learn the reverse chain by leveraging a denoising neural network ${{{{{\boldsymbol{\varepsilon}}_{\boldsymbol{\theta }}} }}}\left( \cdot \right)$, optimized utilizing the objective function \eqref{eq:chmdd}. The proposed GDM takes an initial NF CE ${{{{\mathbf{\hat H}}}^ \star }}$ and a noisy image as input to estimate the noise, and after $T$ refinement steps, it generates the target NF channel.


\textit{2) Gaussian Diffusion Process:} Consider a data sample that is sampled from the actual channel distribution, denoted as ${\mathbf{H}_0} \sim q\left( \mathbf{H} \right)$. For ease of presentation, ${\mathbf{H}}_0$ is adopted to denote ${\mathbf{H}}$ in \eqref{eq:error}. Define a forward process where Gaussian noise is gradually added to the sample in $T$ steps. The level of noise at each step is governed by a variance schedule $\left\{ {{\beta _t} \in \left( {0,1} \right)} \right\}_{t = 1}^T$. Specifically, the forward process is described as \cite{Ho11}
\begin{align}
	\label{eq:marginal ddpm}
	q\left( {{\mathbf{H}_t}\left| {{\mathbf{H}_{t - 1}}} \right.} \right) &= \mathcal{N}\left( {{\mathbf{H}_t};\sqrt {1 - {\beta _t}} {\mathbf{H}_{t - 1}},{\beta _t}\mathbf{I}} \right), \\
	\label{eq:14}
	q\left( {{\mathbf{H}_{1:T}}\left| {{\mathbf{H}_0}} \right.} \right) &= \prod\nolimits_{t = 1}^T {q\left( {{\mathbf{H}_t}\left| {{\mathbf{H}_{t - 1}}} \right.} \right)}. 
\end{align}
Utilizing the reparameterization technique, $\mathbf{H}_t$ can be sampled in a closed form at any given time step $t$:
\begin{align}\label{eq:ht}
	{{{\mathbf{H}}}_t} &= \sqrt {{\alpha _t}} {{{\mathbf{H}}}_{t - 1}} + \sqrt {1 - {\alpha _t}} {\boldsymbol{\varepsilon}_{t - 1}}\nonumber\\
	&= \sqrt {{\alpha _t}} \left(\! {\sqrt {{\alpha _{t - 1}}} {{{\mathbf{H}}}_{t - 2}} \!+\! \sqrt {1 \!-\! {\alpha _{t - 1}}} {\boldsymbol{\varepsilon}_{t - 2}}} \!\right) \!+\! \sqrt {1 \!-\! {\alpha _t}} {\boldsymbol{\varepsilon}_{t - 1}}\nonumber\\
	&=\sqrt {{\alpha _t}{\alpha _{t - 1}}} {{{\mathbf{H}}}_{t - 2}} + \sqrt {{{\sqrt {{\alpha _t} \!-\! {\alpha _t}{\alpha _{t - 1}}} }^2} \!+\! {{\sqrt {1 \!-\! {\alpha _t}} }^2}} {\overline {\boldsymbol{\varepsilon}} _{t - 2}}\nonumber\\
	&=\sqrt {{{\overline \alpha  }_t}} {{{\mathbf{H}}}_0} + \sqrt {1 - {{\overline \alpha  }_t}} \boldsymbol{\varepsilon},
\end{align}
\begin{align}\label{eq:chmd}
	q\left( {{\mathbf{H}_t}\left| {{\mathbf{H}_0}} \right.} \right) = \mathcal{N}\left( {{\mathbf{H}_t};\sqrt {{{\overline \alpha  }_t}} {\mathbf{H}_0},\left(1 - {{\overline \alpha  }_t}\right)  \mathbf{I}} \right),
\end{align}
where ${\alpha _t} = 1 - {\beta _t}$, ${\overline \alpha  _t} = \prod\nolimits_{i = 1}^t {{\alpha_i}}$, ${\boldsymbol{\varepsilon}_{t - 1}},{\boldsymbol{\varepsilon}_{t - 2}, {\overline {\boldsymbol{\varepsilon}} _{t - 2}},...,{\boldsymbol{\varepsilon}}} \sim \mathcal{N}\left( {\mathbf{0},\mathbf{I}} \right)$, and ${\overline {\boldsymbol{\varepsilon}} _{t - 2}}$ merges two Gaussian noises. Usually, we set ${\beta _1} < {\beta _2} < ... < {\beta _T}$. When ${{\bar \alpha }_T}\! \to\! 0$, $q\left( {{{\mathbf{H}}_T}\left| {{{\mathbf{H}}_0}} \right.} \right) \approx \mathcal{N}\left( {{{\mathbf{H}}_T};\mathbf{0},{\mathbf{I}}} \right)$, as $T \!\to\! \infty$.

\textit{3) Noise-Oriented Iterative Refinement:} 
For traditional GDMs, the reverse process is modeled as a backward Markov chain, starting from pure Gaussian noise ${{\mathbf{H}}_T} \sim \mathcal{N}\left( {\mathbf{0},{\mathbf{I}}} \right)$. \textit{To refine the NF CE, we introduce ${{{{\mathbf{\hat H}}}^ \star }}$ as side information to guide the generation process of the backward Markov chain.} Specifically, in the reverse process, we design a sampling-friendly distribution 	${p_{\boldsymbol{\theta }}}\left( {{{\mathbf{H}}_{0:T}}}\left|{{{{\mathbf{\hat H}}}^ \star }}\right. \right)$ \textit{conditioned }on ${{{{\mathbf{\hat H}}}^ \star }}$, which is represented as
\begin{align}	\label{eq:25}
	{p_{\boldsymbol{\theta }}}\left( {{{\mathbf{H}}_{0:T}}}\left|{{{{\mathbf{\hat H}}}^ \star }}\right. \right) = {p}\left( {{{\mathbf{H}}_T}} \right)\prod\nolimits_{t = 1}^T {{p_{\boldsymbol{\theta }}}\left( {{{\mathbf{H}}_{t - 1}}\left| {{{\mathbf{H}}_t}} \right.}, {{{{\mathbf{\hat H}}}^ \star }}\right)}.
\end{align}
The reverse inference process is formulated using isotropic Gaussian conditional distributions, represented as ${p_{{\boldsymbol{\theta}}} }\left( {{{\mathbf{H}}_{t - 1}}\left| {{{\mathbf{H}}_t}} \right.},{{{{\mathbf{\hat H}}}^ \star }} \right)$, which are learned from the training data. Design a Gaussian transition density, which can be learned by a denoising neural network, and is given by
\begin{align} \label{eq:chmd}
\!\!\!\!{p_{{\boldsymbol{\theta}}} }\!\!\left(\! {{{\mathbf{H}}_{t\! -\! 1}}\!\left| {{{\mathbf{H}}_t}} \right.}\!,\!{{{{\mathbf{\hat H}}}^ \star }} \!\right) \!\!=\!\! \mathcal{N}\!\left(\! {{{\mathbf{H}}_{t\! -\! 1}};\!{{\boldsymbol{\mu}}_{{\boldsymbol{\theta}}} }\!\left( \!{{{{\mathbf{\hat H}}}^ \star }}\!,\! {{{\mathbf{H}}_t},\!t} \right)\!\!,\!\mathbf{\Sigma}_{{\boldsymbol{\theta}}} {\!\left(\!{{{{\mathbf{\hat H}}}^ \star }},\! {{{\mathbf{H}}_t},\!t} \!\right)} } \!\!\right)\!\!,\!\!\!
\end{align}
where ${\boldsymbol{\theta}}$ represents the parameters to be learned by the neural network, and ${{\boldsymbol{\mu}}_{{\boldsymbol{\theta}}} }\left( {{{\mathbf{H}}_t},{{{{\mathbf{\hat H}}}^ \star }},t} \right)$ and $\mathbf{\Sigma}_{{\boldsymbol{\theta}}} {\left( {{{\mathbf{H}}_t},{{{{\mathbf{\hat H}}}^ \star }},t} \right)} $ represent the variables ${{\mathbf{H}}_t}$, ${{{{\mathbf{\hat H}}}^ \star }}$, and $t$, with ${\boldsymbol{\theta}}$ being the mean and covariance functions of the parameters.

The key to the reverse process is to train the parameters ${\boldsymbol{\theta}}$ so that the reverse Markov chain, conditioned on ${{{{\mathbf{\hat H}}}^ \star }}$, closely approximates the inverse of the forward Markov chain \eqref{eq:14}. Specifically, this transition density can be expressed in the inverse form of the forward Markov chain \eqref{eq:14} as 
\begin{align}\label{eq:chmd}
	q\left( {{{\mathbf{H}}_{t - 1}}\left| {{{\mathbf{H}}_t},{{\mathbf{H}}_0}} \right.} \right) = \mathcal{N}\left( {{{\mathbf{H}}_{t - 1}};{{{{\tilde {\boldsymbol{\mu}}_t} }}}\left( {{{\mathbf{H}}_t},{{\mathbf{H}}_0}} \right),{{\tilde \beta }_t}{\mathbf{I}}} \right).
\end{align}
Then, we have the explicit expressions for ${{{\tilde {\boldsymbol{\mu}}}}}$, ${\tilde \beta }_t$:
\begin{align}\label{eq:ht-1}
	q\left( {{{\mathbf{H}}_{t - 1}}\left| {{{\mathbf{H}}_t},{{\mathbf{H}}_0}} \right.} \right) = \frac{{q\left( {{{\mathbf{H}}_t}|{{\mathbf{H}}_{t - 1}},{{\mathbf{H}}_0}} \right)q\left( {{{\mathbf{H}}_{t - 1}}|{{\mathbf{H}}_0}} \right)}}{{q\left( {{{\mathbf{H}}_t}|{{\mathbf{H}}_0}} \right)}}.
\end{align}
Leveraging the properties of the Markov chain, \eqref{eq:ht-1} can be further rewritten by \eqref{eq:29}, displayed at the top of this page. Note that $C\left( {{{\mathbf{H}}_t},{{\mathbf{H}}_0}} \right)$ captures functions not involving ${{\mathbf{H}}_{t - 1}}$, with details omitted. Based on \eqref{eq:ht} and \eqref{eq:29}, we have
\begin{figure*}[ht] 
	\begin{align}\label{eq:29}
		q\left( {{{\mathbf{H}}_{t - 1}}\left| {{{\mathbf{H}}_t},{{\mathbf{H}}_0}} \right.} \right)\!
		&= \exp \left( { - \frac{1}{2}\left( {\frac{{{\alpha _t}}}{{{\beta _t}}} + \frac{1}{{1 - {{\bar \alpha }_{t - 1}}}}} \right){\mathbf{H}}_{t - 1}^2 - \left( {\frac{{2\sqrt {{\alpha _t}} }}{{{\beta _t}}}{{\mathbf{H}}_t} + \frac{{2\sqrt {{{\bar \alpha }_t}} }}{{1 - {{\bar \alpha }_t}}}{{\mathbf{H}}_0}} \right){{\mathbf{H}}_{t - 1}} + C\left( {{{\mathbf{H}}_t},{{\mathbf{H}}_0}} \right)} \right)
	\end{align}
    \hrule
\end{figure*}
\begin{align}
	\label{eq:chmd}
	{{\tilde \beta }_t} &= \frac{1}{{\frac{{{\alpha _t}}}{{{\beta _t}}} + \frac{1}{{1 - {{\bar \alpha }_{t - 1}}}}}} = \frac{{1 - {{\bar \alpha }_{t - 1}}}}{{1 - {{\bar \alpha }_t}}}{\beta _t},\\
    \label{eq:uthate1}
	{{{{\tilde {\boldsymbol{\mu}}_t} }}}\left( {{{\mathbf{H}}_t},{{\mathbf{H}}_0}} \right) &= \frac{1}{{\sqrt {{\alpha _t}} }}\left({{\mathbf{H}}_t} - \frac{{{\beta _t}}}{{\sqrt {1 - {{\bar \alpha }_t}} }}{\boldsymbol{\varepsilon}_t}\right).
\end{align}

For the denoising neural network ${{{{{\boldsymbol{\varepsilon}}_{\boldsymbol{\theta }}} }}}\left( \cdot \right)$ to be trained, the likelihood of the observed channel sample ${{{\mathbf{H}}_0}}$ conditioned on ${{{{\mathbf{\hat H}}}^ \star }}$ needs to be solved. Under the variational inference framework, we can derive the evidence lower bound (ELBO) as a proxy objective function to maximize the conditional marginal distribution $\log p\left( {{{\mathbf{H}}_0}}\left|{{{{\mathbf{\hat H}}}^ \star }}\right.  \right)$, i.e.,
\begin{subequations}\label{eq:}
	\begin{align}
		\!\!\log p\left( {{{\mathbf{H}}_0}}\left|{{{{\mathbf{\hat H}}}^ \star }}\right.  \right) &= \log \int {p\left( {{{\mathbf{H}}_{0:T}}}\left|{{{{\mathbf{\hat H}}}^ \star }}\right. \right)} d{{\boldsymbol{{\boldsymbol{\mathbf{H}}}}}_{1:T}}\\ 
		&\mathop  \geqslant \limits^{(a)} \!{\mathbb{E}_{q\left({{\mathbf{H}_{1:T}}\left| {{\mathbf{H}_0}} \right.}  \right)}}\!\!\!\left( {\log \frac{{p\left( {{{\mathbf{H}}_{0:T}}}\left|{{{{\mathbf{\hat H}}}^ \star }}\right.  \right)}}{{q\left( {{\mathbf{H}_{1:T}}\left| {{\mathbf{H}_0}} \right.}  \right)}}} \right)\!, \label{eq:34f}
	\end{align}
\end{subequations}
where $\mathop  \geqslant \limits^{(a)} $ in \eqref{eq:34f} follows from Jensen's inequality. To further derive the ELBO, \eqref{eq:34f} can be rewritten as \eqref{eq:34}, displayed at the top of the next page.
\begin{figure*}[ht] 
	\begin{subequations}\label{eq:}
		\begin{align}\label{eq:343}
			&\!\!\log p\left( {{{\mathbf{H}}_0}}\left|{{{{\mathbf{\hat H}}}^ \star }}\right.  \right) 
			\geqslant {\mathbb{E}_{q\left({{\mathbf{H}_{1:T}}\left| {{\mathbf{H}_0}} \right.}  \right)}}\left( {\log \frac{{p\left( {{{{{\mathbf{H}}}}_T}} \right){p_{{\boldsymbol{\theta}}}}\left( {{{{\mathbf{H}}}_0}\left| {{{{\mathbf{H}}}_1},{ {{{{\mathbf{\hat H}}}^ \star }}} } \right.} \right)}}{{q\left( {{{{\mathbf{H}}}_1}\left| {{{{\mathbf{H}}}_0}} \right.} \right)}} + \log \frac{{q\left( {{{{\mathbf{H}}}_1}\left| {{{\mathbf{H}}_0}} \right.} \right)}}{{q\left( {{{\mathbf{H}}_T}\left| {{{{\mathbf{H}}}_0}} \right.} \right)}} + \log \prod\limits_{t = 2}^T {\frac{{{p_{{\boldsymbol{\theta}}}}\left( {{{{\mathbf{H}}}_{t - 1}}\left| {{{{\mathbf{H}}}_t},{ {{{{\mathbf{\hat H}}}^ \star }}} } \right.} \right)}}{{q\left( {{{{\mathbf{H}}}_{t - 1}}\left| {{{{\mathbf{H}}}_t},{{{\mathbf{H}}}_0}} \right.} \right)}}} } \right)\\
			&\!\!\!= {\mathbb{E}_{q\left( {{{{\mathbf{H}}}_1}\left| {{{{\mathbf{H}}}_0}} \right.} \right)}}\!\left(\! {\log {p_{{\boldsymbol{\theta}}}}\!\left(\! {{{{\mathbf{H}}}_0}\!\left| {{{{\mathbf{H}}}_1},{ {{{{\mathbf{\hat H}}}^ \star }}} } \right.} \!\right)} \!\right) \!+\! {\mathbb{E}_{q\left( {{{{\mathbf{H}}}_T}\left| {{{{\mathbf{H}}}_0}} \right.} \right)}}\!\!\left(\! {\log \frac{{p\left( {{{{\mathbf{H}}}_T}} \right)}}{{q\left( {{{{\mathbf{H}}}_T}\!\left| {{{{\mathbf{H}}}_0}} \right.} \right)}}} \!\right) \!+ \sum\limits_{t = 2}^T {{\mathbb{E}_{q\left( {{{{\mathbf{H}}}_t},{{{\mathbf{H}}}_{t - 1}}\left| {{{{\mathbf{H}}}_0}} \right.} \right)}}\left( {\log \frac{{{p_{\boldsymbol{\theta }}}\left( {{{{\mathbf{H}}}_{t - 1}}\left| {{{{\mathbf{H}}}_t},{ {{{{{\mathbf{\hat H}}}^ \star }}}} } \right.} \right)}}{{q\left( {{{{\mathbf{H}}}_{t - 1}}\left| {{{{\mathbf{H}}}_t},{{{\mathbf{H}}}_0}} \right.} \right)}}} \right)}  \\
			&\!\!= \underbrace {{\mathbb{E}_{q\left( {{{{\mathbf{H}}}_1}\!\left| {{{{\mathbf{H}}}_0}} \right.} \!\right)}}\!\!\left(\! {\log\! {p_{{\boldsymbol{\theta}}}}\!\!\left(\! {{{{\mathbf{H}}}_0}\!\left| {{{{\mathbf{H}}}_1}\!,\!{ {{{{\mathbf{\hat H}}}^ \star }}} } \right.} \!\!\right)} \!\!\right)}_{{\mathcal{L}_{{a}}}}\!\!-\!\! \underbrace {\sum\limits_{t = 2}^T\! {{\mathbb{E}_{q\left( {{{{\mathbf{H}}}_t}\!\left| {{{{\mathbf{H}}}_0}} \right.} \!\right)}}\!\!\left(\!\! {{{{D_\text{KL}}}}\!\!\left(\!\! {\left. {q\!\left( {{{{\mathbf{H}}}_{t \!-\! 1}}\!\left| {{{{\mathbf{H}}}_t},\!{{{\mathbf{H}}}_0}} \right.}\! \!\right)\!} \right\|\!{p_{\boldsymbol{\theta }}}\!\left(\! {{{{\mathbf{H}}}_{t\! - \!1}}\!\left| {{{{\mathbf{H}}}_t},\!{ {{{{\mathbf{\hat H}}}^ \star }}} } \right.} \!\!\right)} \!\!\right)} \!\!\right)} }_{{\mathcal{L}_{{b}}}} \!\!-\!\! \underbrace {{{D_\text{KL}}}\!\!\left(\! {\left. {q\!\left(\! {{{{\mathbf{H}}}_T}\!\left| {{{{\mathbf{H}}}_0}} \right.}\! \right)} \right\|\!p\!\left(\! {{{{\mathbf{H}}}_T}} \!\right)} \!\right)}_{{\mathcal{L}_{{c}}}}\!=\! \mathcal{L}_{\rm{ELBO}}\!\left( {\boldsymbol{\theta}}  \right) \label{eq:34}
		\end{align}
	\end{subequations}
	\hrule
\end{figure*}
Then, the parameter $\boldsymbol{\theta}$ can be learned by optimizing the negative ELBO:
\begin{align}\label{eq:27}
	\mathop {{\text{argmin}}}\limits_{\boldsymbol{\theta}} 
	\mathcal{L}\left( {{\boldsymbol{\theta}}} \right) = {\mathbb{E} }\left( { - {\mathcal{L}_{{\text{ELBO}}}}\left( {{\boldsymbol{\theta}}} \right)} \right).
\end{align}
Note that ${{{\mathcal{L}}_c}}$ is constant, and ${{{\mathcal{L}}_a}}$ can be approximated using a Monte Carlo estimate. Then, the bound of the training objective is dominated by ${{{\mathcal{L}}_b}}$. Consequently, it is essential to train a ${{{{{\boldsymbol{\varepsilon}}_{\boldsymbol{\theta }}} }}}\left( \cdot \right)$ to fit the conditioned probability distributions in the reverse process, i.e., to make the KL divergence of ${p_{\boldsymbol{\theta }}}\left( {{{\mathbf{H}}_{t - 1}}\left| {{{{\mathbf{H}}_t}},{{{{\mathbf{\hat H}}}^ \star }}} \right.} \right)$ and $q\left( {{{\mathbf{H}}_{t - 1}}\left| {{{\mathbf{H}}_t},{{\mathbf{H}}_0}} \right.} \right)$ as small as possible. Since the variance ${\bf{\Sigma}}_{\boldsymbol{\theta }} {\left({{{{\mathbf{\hat H}}}^ \star }}, {{{\mathbf{H}}_t},t} \right)} $ is usually set to a constant ${{\tilde \beta }_t}$ \cite{Ho11}, the learnable parameters are only in the mean ${{\boldsymbol{\mu}_{\boldsymbol{\theta }}}}$. Therefore, we only need to enable the GDM conditioned on ${{{{\mathbf{\hat H}}}^ \star }}$ to predict ${{\tilde {\boldsymbol{\mu}}}_t} = {{\left( {{{\mathbf{H}}_t} - {{\boldsymbol{\varepsilon }}_t}{{\left( {1 - {\alpha _t}} \right)} \mathord{\left/
					{\vphantom {{\left( {1 - {\alpha _t}} \right)} {\sqrt {1 - {{\bar \alpha }_t}} }}} \right.
					\kern-\nulldelimiterspace} {\sqrt {1 - {{\bar \alpha }_t}} }}} \right)} \mathord{\left/
		{\vphantom {{\left( {{{\mathbf{H}}_t} - {{\boldsymbol{\varepsilon }}_t}{{\left( {1 - {\alpha _t}} \right)} \mathord{\left/
								{\vphantom {{\left( {1 - {\alpha _t}} \right)} {\sqrt {1 - {{\bar \alpha }_t}} }}} \right.
								\kern-\nulldelimiterspace} {\sqrt {1 - {{\bar \alpha }_t}} }}} \right)} {\sqrt {{\alpha _t}} }}} \right.
		\kern-\nulldelimiterspace} {\sqrt {{\alpha _t}} }}$:
\begin{align}\label{eq:uthate}
	{{\boldsymbol{\mu}}_{\boldsymbol{\theta }}}\left({{{{\mathbf{\hat H}}}^ \star }}, {{{\mathbf{H}}_t},t} \right) \!=\! \frac{1}{{\sqrt {{\alpha _t}} }}\!\left(\! {{{\mathbf{H}}_t} \!-\! \frac{{1 \!-\! {\alpha _t}}}{{\sqrt {1 \!-\! {{\bar \alpha }_t}} }}{{\boldsymbol{\varepsilon}}_{\boldsymbol{\theta }}}\!\!\left(\!{{{{\mathbf{\hat H}}}^ \star }}, {{{\mathbf{H}}_t},t} \!\right)} \!\right),
\end{align}
\begin{align}\label{eq:}
	{{\mathbf{H}}_{t \!-\! 1}} \!\!=\!\! \mathcal{N}\!\!\left(\!\! {{\mathbf{H}}_{t \!-\! 1}}\!;\!\!\frac{1}{{\sqrt {{\alpha _t}} }}\!\!\left(\! {{{\mathbf{H}}_t} \!\!-\!\! \frac{{1 \!-\! {\alpha _t}}}{{\sqrt {1 \!-\! {{\bar \alpha }_t}} }}{{\boldsymbol{\varepsilon}}_{\boldsymbol{\theta }}}\!\!\left(\!{{{{\mathbf{\hat H}}}^ \star }}\!, {{{\mathbf{H}}_t},t} \right)} \right),{{\tilde \beta }_t} {\mathbf{I}}  \! \!\right)\!.
\end{align}
Substituting \eqref{eq:uthate1}, and \eqref{eq:uthate} into ${{{\mathcal{L}}_b}}$ in \eqref{eq:34}, yields
\begin{align}\label{eq:chmd}
	{{{\mathcal{L}}_b}} 
    = \prod\limits_{t = 2}^T  {\mathbb{E}_{{{\mathbf{H}}_0},{\boldsymbol{\varepsilon}}_t}}\bigg(\! \mathfrak{B}\big\| {{\boldsymbol{\varepsilon}}_t}  -{{\boldsymbol{\varepsilon}}_{\boldsymbol{\theta }}}\left({{{{\mathbf{\hat H}}}^ \star }}, {{{\mathbf{H}}_t},t} \right) \big\|_2^2 \!\bigg),
\end{align}
where $\mathfrak{B}=\frac{{{{\left( {1 - {\alpha _t}} \right)}^2}}}{{2{\alpha _t}\left( {1 - {{\bar \alpha }_t}} \right)\tilde \beta _t^2}}$. With \eqref{eq:chmd}, the objective $\mathcal{L}(\boldsymbol{\theta})$ in \eqref{eq:27} can be simplified to:
\begin{align}\label{eq:chmdd}
 {{{\mathcal{L}}}_{\rm{Simp}}}(\boldsymbol{\theta}) \!&:=\!\!\! \sum\limits_{t = 1}^T\!{\mathbb{E}_{{{\mathbf{H}}_0},{{\boldsymbol{\varepsilon }}_t} }}\!\Big( \big\| {{\boldsymbol{\varepsilon }}_t} - {{\boldsymbol{\varepsilon}}_{\boldsymbol{\theta }}}\left({{{{\mathbf{\hat H}}}^ \star }}, {{{\mathbf{H}}_t},t} \right) \big\|_2^2 \Big)\!.
\end{align} 
\begin{algorithm}[ht]
	\caption{Training Scheme for the Denoising Neural Network Conditioned on the Initial CE}
	\label{alg:train}
	\begin{algorithmic}[1]
		\REPEAT
		\STATE
		Load the training channel pairs $\left(\!{ {{{{\mathbf{\hat H}}}^ \star }},{{\mathbf{H}}_0}} \!\right) \!\sim\! p\left(\!{ {{{{\mathbf{\hat H}}}^ \star }},{{\mathbf{H}}_0}} \!\right)$\STATE
		Obtain time steps $t \sim {\rm{Uniform}}({1,...,T})$
		\STATE
		Randomly generate a noise tensor with the same dimensions as ${{\mathbf{H}}_0}$, ${{\boldsymbol{\varepsilon }}_{t}} \sim \mathcal{N}\left( {\mathbf{0},\mathbf{I}} \right)$
		\STATE
		Add noise gradually to the target NF channel ${{\mathbf{H}}_0}$ according to \eqref{eq:ht} to execute the diffusion process.
		\STATE Feed the corrupted NF channel ${{\mathbf{H}}_t}$, the initial CE ${{{{\mathbf{\hat H}}}^ \star }}$, and time step $t$ into the denoising neural network ${{{{{\boldsymbol{\varepsilon}}_{\boldsymbol{\theta }}} }}}\left( \cdot \right)$\STATE
		Take gradient descent step on the optimization objective \eqref{eq:chmdd} to update the network parameters ${\boldsymbol{\theta }}$:\\
		\quad \quad \quad $ \nabla_{\boldsymbol{\theta }} \big\| {{\boldsymbol{\varepsilon }}_t} - {{\boldsymbol{\varepsilon}}_{\boldsymbol{\theta }}}\left({{{{\mathbf{\hat H}}}^ \star }}, {{{\mathbf{H}}_t},t} \right) \big\|_2^2 $
		\UNTIL the optimization objective \eqref{eq:chmdd} converges
	\end{algorithmic}
\end{algorithm}
\begin{algorithm}[ht]
	\caption{Inferring the Target NF Channel in the Reverse Process, Conditioned on the Initial CE, Through $T$ Iterative Refinement Steps.}
	\label{alg:infer}
	\begin{algorithmic}[1]
		\STATE
		Load the pretrained model ${{{{{\boldsymbol{\varepsilon}}_{\boldsymbol{\theta }}} }}}\left( \cdot \right)$ and its weights ${\boldsymbol{\theta }}$
		\STATE
		Obtain the entirely corrupted NF channel ${{\mathbf{H}}_T}\sim \mathcal{N}\left( {\mathbf{0},\mathbf{I}} \right)$, and the initial CE ${{{{\mathbf{\hat H}}}^ \star }}$
		\FOR{$t=T,..,1$}
		\STATE
		${\boldsymbol{\varepsilon}^*} \sim\mathcal{N}\left( {\mathbf{0},\mathbf{I}} \right)$ if $t>1$, else ${\boldsymbol{\varepsilon}^*} =0$
		\STATE
		Execute the refinement step according to \eqref{eq:3334}:\\
		\quad ${{{{\mathbf{H}}}_{t\!-\!1}}} \!\leftarrow\! \frac{1}{{\sqrt {{\alpha _t}} }}\!\!\left(\!\! {{{{{\mathbf{H}}}}_t} \!\!-\!\! \frac{{1 \!-\! {\alpha _t}}}{{\sqrt {1 \!-\! {{\bar \alpha }_t}} }}{{\boldsymbol{\varepsilon}}_{\boldsymbol{\theta }}}\!\left(\!{{ {{{{\mathbf{\hat H}}}^ \star }}}, {{{\mathbf{H}}}_t},t} \!\right)} \!\!\right) \!+\sqrt{ \tilde \beta} {\boldsymbol{\varepsilon}^*}$
		\ENDFOR
		\RETURN Refined NF channel ${{{{\hat{{\mathbf{H}}}_0}}}}$
	\end{algorithmic}
\end{algorithm}

By optimizing this objective in \eqref{eq:chmdd}, the trained GDM, guided by the side information ${{{{\mathbf{\hat H}}}^ \star }}$, predicts the noise ${{\boldsymbol{\varepsilon}}_t}$ and approximates the target NF channel through \eqref{eq:ht}, i.e., 
\begin{align}\label{eq:33}
	{{{{\hat{{\mathbf{H}}}_0}}}} \!\!=\!\! \frac{1}{{\sqrt {{{\bar \alpha }_t}} }}\!\!\left(\!\! {{{{{\mathbf{H}}}_t}} \!\!-\!\! \sqrt {1 \!\!-\! {{\bar \alpha }_t}} {{\boldsymbol{\varepsilon }}_{\boldsymbol{\theta }}}\big({{{{\mathbf{\hat H}}}^ \star }}, \underbrace{\sqrt {{{\bar \alpha }_t}} {{{{\mathbf{H}}}_0}} \!+\! \sqrt {1 \!-\! {{\bar \alpha }_t}} {{\boldsymbol{\varepsilon }}}}_{{{{\mathbf{H}}}_t}},t \big)} \!\!\right)\!.
\end{align}
Through reparameterization, \eqref{eq:33} represents the results of the iterative refinement of the initial NF CE, with each iteration of the side-information-guided GDM represented by
\begin{align}\label{eq:3334}
	\!\!\!\!\!{{{{\mathbf{H}}}_{t\!-\!1}}} \!\leftarrow\! \frac{1}{{\sqrt {{\alpha _t}} }}\!\!\left(\!\! {{{{{\mathbf{H}}}}_t} \!\!-\!\! \frac{{1 \!-\! {\alpha _t}}}{{\sqrt {1 \!-\! {{\bar \alpha }_t}} }}{{\boldsymbol{\varepsilon}}_{\boldsymbol{\theta }}}\!\left(\!{{ {{{{\mathbf{\hat H}}}^ \star }}}, {{{\mathbf{H}}}_t},t} \!\right)} \!\!\right) \!+\sqrt{ \tilde \beta_t} {\boldsymbol{\varepsilon}^*},\!\!\!\!\!\!
\end{align}
where ${\boldsymbol{\varepsilon}^*} \sim\mathcal{N}\left( {\mathbf{0},\mathbf{I}} \right)$. Note that the noise estimation step in \eqref{eq:3334} is analogous to a step in Langevin dynamics within score-based generative models \cite{song2019generative}, which is equivalent to the estimation of the first derivative of the log-likelihood of the observed samples, also referred to as the Stein score. For clarity, the offline training strategy and the online iterative refinement procedure of the proposed GDM, conditioned on the initial CE, are summarized in \textbf{Algorithm \ref{alg:train}} and \textbf{Algorithm \ref{alg:infer}}, respectively.

\subsection{Accelerating GDM for CE Refinement}

Although the GDM conditioned on side information can achieve high-quality generation outcomes without the need for adversarial training, it requires simulating a Markov chain for numerous steps to generate the target channel. From a variational perspective, when $T$ is larger, the reverse process is closer to a Gaussian, so that the generative process modeled with Gaussian conditional distribution becomes a good approximation. However, a relatively large $T$ would cause the GDM to sample too slowly, making it impractical for some tasks where computation is limited and latency is critical. To further improve the proposed GDM, we introduce a non-Markovian GDM (NM-GDM) to accelerate the sampling process. Specifically, the key part in \eqref{eq:34} is ${{D_{\rm{KL}}}}\left( {q\left( {{{\mathbf{H}}_{t - 1}}\left| {{{\mathbf{H}}_t},{{\mathbf{H}}_0}} \right.} \right)||{p_{\boldsymbol{\theta }}}\left( {{{{\mathbf{H}}}_{t - 1}}\left| {{{{\mathbf{H}}}_t},{ {{{{\mathbf{\hat H}}}^ \star }}} } \right.} \right)} \right)$, which means that the GDM is to train ${{p_{\boldsymbol{\theta }}}\left(  {{{{\mathbf{H}}}_{t - 1}}\left| {{{{\mathbf{H}}}_t},{ {{{{\mathbf{\hat H}}}^ \star }}} } \right.} \right)}$ as an approximate solution to ${q\left( {{{\mathbf{H}}_{t - 1}}\left| {{{\mathbf{H}}_t},{{\mathbf{H}}_0}} \right.} \right)}$. Note that the condition variables in ${q\left( {{{\mathbf{H}}_{t - 1}}\left| {{{\mathbf{H}}_t},{{\mathbf{H}}_0}} \right.} \right)}$ are the observed variable ${{{\mathbf{H}}_0}}$ and the latent variable ${{{\mathbf{H}}_t}}$, where $q\left( {{{\mathbf{H}}_t}\left| {{{\mathbf{H}}_0}} \right.} \right)$ is obtained by marginalization of $q\left( {{{\mathbf{H}}_{1:t}}\left| {{{\mathbf{H}}_0}} \right.} \right)$, denoted by
\begin{align}\label{eq:39}
	q\left( {{{\mathbf{H}}_t}\left| {{{\mathbf{H}}_0}} \right.} \right) = \int {q\left( {{{\mathbf{H}}_{1:t}}\left| {{{\mathbf{H}}_0}} \right.} \right)d} {{\mathbf{H}}_{1:t - 1}},
\end{align}
where the joint probability ${q\left( {{{\mathbf{H}}_{1:t}}\left| {{{\mathbf{H}}_0}} \right.} \right)}$ is decomposed in the original GDM in the form of a Markov chain, given by
\begin{align}\label{eq：40}
	q\left( {{{\mathbf{H}}_t}\left| {{{\mathbf{H}}_0}} \right.} \right) = \int {\prod\nolimits_{i = 1}^t {q\left( {{{\mathbf{H}}_i}\left| {{{\mathbf{H}}_{i - 1}}} \right.} \right)d{{\mathbf{H}}_{1:t - 1}}} }.
\end{align}
Note that \eqref{eq：40} is obtained by Markov chain decomposition of ${q\left( {{{\mathbf{H}}_{1:t}}\left| {{{\mathbf{H}}_0}} \right.} \right)}$ in \eqref{eq:39}, and in fact the way of decomposition does not affect the numerical results of \eqref{eq:39}. From a probabilistic perspective, regardless of how $q\left( {{{\mathbf{H}}_{1:t}}\left| {{{\mathbf{H}}_0}} \right.} \right)$ is decomposed, it will ultimately be eliminated through integration. In fact, there exist various valid decomposition strategies, yet none affect the result after integration. In other words, the way $q\left( {{{\mathbf{H}}_{1:t}}\left| {{{\mathbf{H}}_0}} \right.} \right)$ is decomposed does not impact $q\left( {{{\mathbf{H}}_t}\left| {{{\mathbf{H}}_0}} \right.} \right)$. Therefore, an NM-GDM can be obtained by abandoning the Markov chain structure in the forward diffusion process. A non-Markovian form equivalent to the GDM can be obtained by ensuring that the analytical expressions of $q\left( {{{\mathbf{H}}_t}\left| {{{\mathbf{H}}_0}} \right.} \right)$ and ${q\left( {{{\mathbf{H}}_{t - 1}}\left| {{{\mathbf{H}}_t},{{\mathbf{H}}_0}} \right.} \right)}$ are consistent with the original GDM.

Specifically, let $\mathcal{Q}$ denote a family of inference distributions, indexed by a real vector $\sigma  \in \mathbb{R}_{ \geqslant 0}^T$, and ${\sigma ^2}$ denotes the variance of ${q_\sigma }\left( {{{\mathbf{H}}_{t - 1}}\left| {{{\mathbf{H}}_t},{{\mathbf{H}}_0}} \right.} \right)$. Then, the decomposition of $q\left( {{{\mathbf{H}}_{1:T}}\left| {{{\mathbf{H}}_0}} \right.} \right)$ is redefined as
\begin{align}\label{eq：41}
	{q_\sigma }\left( {{{\mathbf{H}}_{1:T}}\left| {{{\mathbf{H}}_0}} \right.} \right) = {q_\sigma }\left( {{{\mathbf{H}}_T}\left| {{{\mathbf{H}}_0}} \right.} \right)\prod\limits_{t = 2}^T {{q_\sigma }\left( {{{\mathbf{H}}_{t - 1}}\left| {{{\mathbf{H}}_t},{{\mathbf{H}}_0}} \right.} \right)},
\end{align}
where ${q_\sigma }\left( {{{\mathbf{H}}_T}\left| {{{\mathbf{H}}_0}} \right.} \right) \sim \mathcal{N}\left( {{{\mathbf{H}}_T};\sqrt {{{\bar \alpha }_T}} {{\mathbf{H}}_0},\left( {1 - {{\bar \alpha }_T}} \right){\mathbf{I}}} \right)$. For any $t > 1$, ${{q_\sigma }\left( {{{\mathbf{H}}_{t - 1}}\left| {{{\mathbf{H}}_t},{{\mathbf{H}}_0}} \right.} \right)}$ is denoted as
\begin{align}\label{eq：42}
	&{q_\sigma }\left( {{{\mathbf{H}}_{t - 1}}\left| {{{\mathbf{H}}_t},{{\mathbf{H}}_0}} \right.} \right) \sim \mathcal{N}\bigg( {{\mathbf{H}}_{t - 1}};{\boldsymbol{\mu}}_\sigma,\sigma _t^2{\mathbf{I}} \bigg),
\end{align}
where ${\boldsymbol{\mu}}_\sigma=\sqrt {{{\bar \alpha }_{t - 1}}} {{\mathbf{H}}_0} +\sqrt {1 - {{\bar \alpha }_{t - 1}} - \sigma _t^2}  \cdot \frac{{{{\mathbf{H}}_t} - \sqrt {{{\bar \alpha }_t}} {{\mathbf{H}}_0}}}{{\sqrt {1 - {{\bar \alpha }_t}} }}$. Note that the mean value in \eqref{eq：42} is defined as a combined function depending on ${{\mathbf{H}}_0}$ and ${{\mathbf{H}}_t}$ to ensure that ${q_\sigma }\left( {{{\mathbf{H}}_T}\left| {{{\mathbf{H}}_0}} \right.} \right) \sim \mathcal{N}\left( {{{\mathbf{H}}_T};\sqrt {{{\bar \alpha }_T}} {{\mathbf{H}}_0},\left( {1 - {{\bar \alpha }_T}} \right){\mathbf{I}}} \right)$ holds, $\forall t>1$. Based on \eqref{eq:ht}, the estimated result ${\widetilde {\mathbf{H}}_0}$ is given by
\begin{align}\label{eq：h0pie}
	{\widetilde {\mathbf{H}}_0} \!\!=\!\! f_{\boldsymbol{\theta}} ^{\left( t \right)}\left({{{{\mathbf{\hat H}}}^ \star }}, {{{\mathbf{H}}_t}} \right): = \frac{{{{\mathbf{H}}_t} - \sqrt {1 - {{\bar \alpha }_t}} {{\boldsymbol{\varepsilon }}_{\boldsymbol{\theta }}}\left({{{{\mathbf{\hat H}}}^ \star }}, {{{\mathbf{H}}_t},t} \right)}}{{\sqrt {{{\bar \alpha }_t}} }},
\end{align}
where ${{{\boldsymbol{\varepsilon}}_{\boldsymbol{\theta }}}\left({{{{\mathbf{\hat H}}}^ \star }}, {{{\mathbf{H}}_t},t} \right)}$ is the noise predicted by the GDM \textit{conditioned} on the side information. Combining \eqref{eq：42} and \eqref{eq：h0pie}, the approximate distribution of ${q_\sigma }\left( {{{\mathbf{H}}_{t - 1}}\left| {{{\mathbf{H}}_t},{{\mathbf{H}}_0}} \right.} \right)$ is given by
\begin{align}\label{eq：43}
	\!{p_{{\boldsymbol{\theta}} ,\sigma }}\!\!\left(\! {{{\mathbf{H}}_{t \!-\! 1}}\!\!\left| {{{{\mathbf{H}}_t}},\!{{{{\mathbf{\hat H}}}^ \star }}} \right.} \!\!\right)\!\!\sim\!\! \mathcal{N}\!\bigg(\!\! {{\mathbf{H}}_{t \!-\! 1}}\!;\!{\widetilde {{\boldsymbol{\mu}}}}_\sigma,\!\sigma _t^2{\mathbf{I}} \!\!\bigg)\!\!\approx\! {q_\sigma }\!\!\left({{{\mathbf{H}}_{t \!-\! 1}}\!\!\left| {{{\mathbf{H}}_t},\!{{\mathbf{H}}_0}} \right.} \!\right)\!,\!\!
\end{align}
where ${\widetilde {{\boldsymbol{\mu}}}}_\sigma=\sqrt {{{\bar \alpha }_{t - 1}}} {{\widetilde {\mathbf{H}}}_0} +\sqrt {1 - {{\bar \alpha }_{t - 1}} - \sigma _t^2}  \cdot \frac{{{{\mathbf{H}}_t} - \sqrt {{{\bar \alpha }_t}} {{\widetilde {\mathbf{H}}}_0}}}{{\sqrt {1 - {{\bar \alpha }_t}} }}$. 
\begin{figure*}[!t]
	\centering
	\includegraphics[scale=0.1096525]{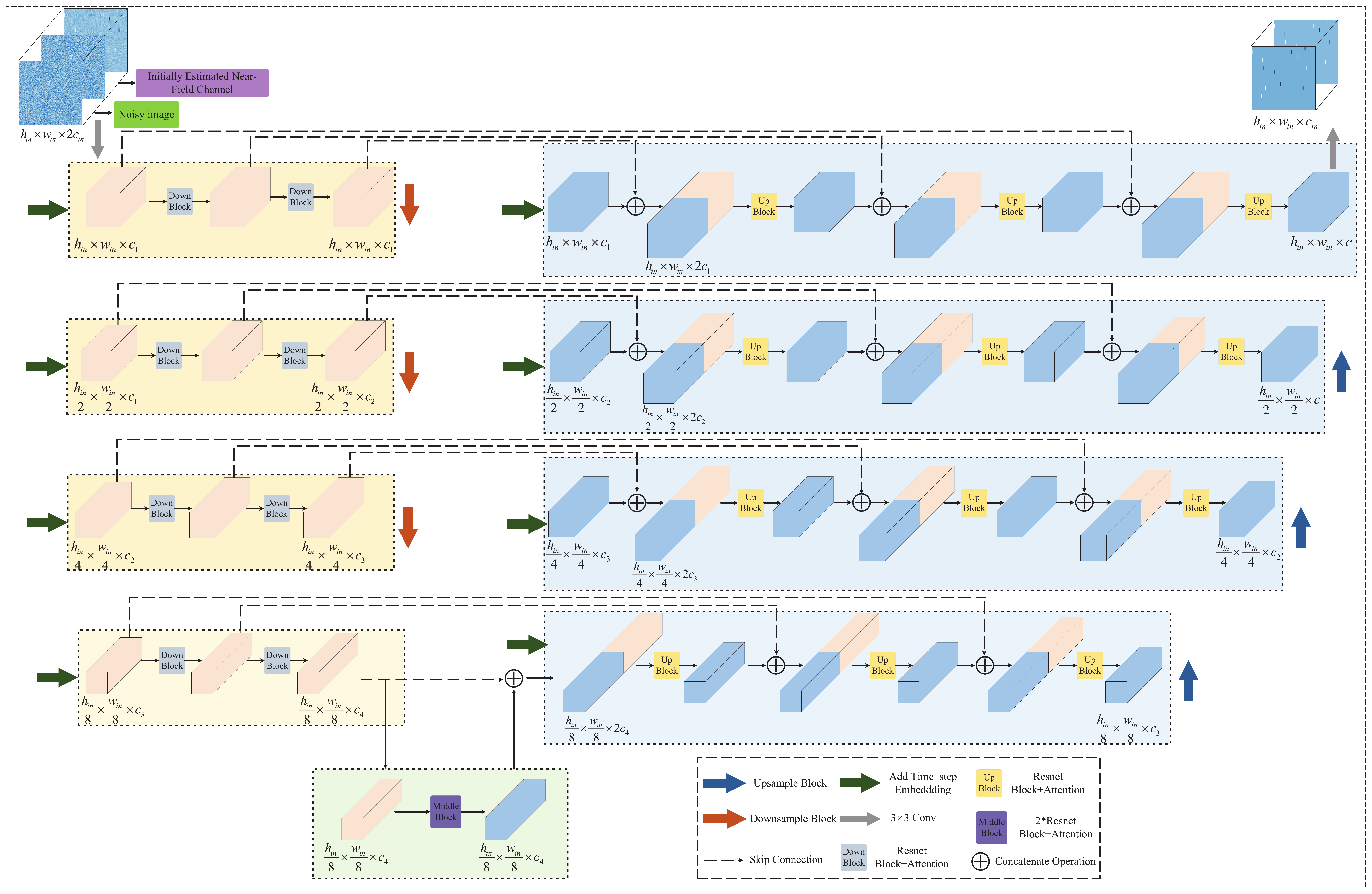}
	\captionsetup{font=footnotesize}
	\caption{Network architecture of the proposed NM-GDM conditioned on the initial CE. In the training experimental setup, the preliminary NF CE dimensions are reshaped to $h_{in} \times w_{in}$ (e.g., $256 \times 256$), with the number of channels $c_{in}$ set to 2, and concatenated with the noisy image of the same dimensions. Within the same level, the height $h$ and width $w$ of the feature maps remain unchanged, while the number of channels $c$ is controlled by the channel number multiplier ${\eta _c} = c_1:c_2:c_3:c_4$. Its configuration, along with the base channels $c_1$ (e.g., 8, 16, 32, 64) at the first level in the latent space, is thoroughly compared and discussed in Section V.B.}
	\label{fig:model}
\end{figure*}

For the NM-GDM \textit{conditioned} on ${{{{\mathbf{\hat H}}}^ \star }}$, we then redefine the reverse process with a fixed prior ${p_{\boldsymbol{\theta}} }\left( {{{\mathbf{H}}_T}} \right) = \mathcal{N}\left( {{{\mathbf{H}}_T};\mathbf{0},{\mathbf{I}}} \right)$ and
\begin{align}\label{eq：43}
    p_{\boldsymbol{\theta}} ^{\left( t \right)}\!\!\left(\! {{{\mathbf{H}}_{t \!-\! 1}}\!\!\left|\! {{{{\mathbf{H}}_t}},\!{{{{\mathbf{\hat H}}}^ \star }}} \right.}\!\! \right) \!\!=\!\! \left\{\!\!\!\!\! {\begin{array}{*{20}{c}}
    		{\!\!\!\!\mathcal{N}\left( {f_{\boldsymbol{\theta}} ^{\left( 1 \right)}\left({{{{\mathbf{\hat H}}}^ \star }}, {{{\mathbf{H}}_1}} \right),\sigma _1^2{\mathbf{I}}} \right),   \text{if}{\text{ }}t = 1} \\ 
    		{{q_\sigma }\!\!\left(\! {{{\mathbf{H}}_{t \!-\! 1}}\!\!\left|\! {{{\mathbf{H}}_t},\!f_{\boldsymbol{\theta}} ^{\left( t \right)}\!\!\left(\!{{{{\mathbf{\hat H}}}^ \star }}, {{{\mathbf{H}}_t}} \!\right)} \right.} \!\right),  \text{otherwise}} 
    \end{array}} \right.\!\!\!\!,
\end{align}
where ${{{\mathbf{H}}_{t - 1}}}$ is given by
\begin{align}\label{eq：46}
	{{\mathbf{H}}_{t - 1}} &\!=\! \sqrt {{{\bar \alpha }_{t\! -\! 1}}} {\widetilde {\mathbf{H}}_0} \!+\! \sqrt {1 \!-\! {{\bar \alpha }_{t \!-\! 1}} \!-\! \sigma _t^2}  \cdot \frac{{{{\mathbf{H}}_t} \!-\! \sqrt {{{\bar \alpha }_t}} {{\widetilde {\mathbf{H}}}_0}}}{{\sqrt {1 - {{\bar \alpha }_t}} }} + {\sigma _t}{{\boldsymbol{\varepsilon }}_t}\nonumber\\
	&= {{\dot{{\boldsymbol{\varepsilon }}}_{{\boldsymbol{\theta }},t}}} + {\sigma _t}{{\boldsymbol{\varepsilon }}_t},
\end{align}
and ${{\dot{{\boldsymbol{\varepsilon }}}_{{\boldsymbol{\theta }},t}}}\!\!=\!\! \sqrt {{{\bar \alpha }_{t \!-\! 1}}} \!\left(\! {\frac{{{{\mathbf{H}}_t} - \sqrt {1 - {{\bar \alpha }_t}} {{\boldsymbol{\varepsilon }}_{\boldsymbol{\theta }}}\left( {{{{\mathbf{\hat H}}}^ \star }},{{{\mathbf{H}}_t},t} \right)}}{{\sqrt {{{\bar \alpha }_t}} }}} \!\right) \!+\!\sqrt {1 \!-\! {{\bar \alpha }_{t \!-\! 1}} \!-\! \sigma _t^2}  \cdot {{\boldsymbol{\varepsilon }}_{\boldsymbol{\theta }}}\left({{{{\mathbf{\hat H}}}^ \star }}, {{{\mathbf{H}}_t},t} \right)$.
Combining \eqref{eq：41} and \eqref{eq:25}, this variational objective in the NM-GDM \textit{conditioned} on ${{{{\mathbf{\hat H}}}^ \star }}$ is written as
\begin{align}\label{eq：47}
	\mathcal{L}(\boldsymbol{\theta}) &\!\!=\!\! {\mathbb{E}_{{{\mathbf{H}}_{0:T}} \sim {q_\sigma }\!\left( {{{\mathbf{H}}_{0:T}}} \!\right)}}\!\Big[\! {\log\! {q_\sigma }\!\left( {{{\mathbf{H}}_{1:T}}\!\left| {{{\mathbf{H}}_0}} \right.} \!\right) \!\!-\!\! \log\! {p_{\boldsymbol{\theta }}}\!\!\left(\! {{{\mathbf{H}}_{0:T}}}\!\left|{{{{\mathbf{\hat H}}}^ \star }} \right. \right)} \Big]\nonumber\\
	&\!\!={\mathbb{E}_{{{\mathbf{H}}_{0:T}} \sim {q_\sigma }\left( {{{\mathbf{H}}_{0:T}}} \right)}}\Big[\log {q_\sigma }\left( {{{\mathbf{H}}_T}\left| {{{\mathbf{H}}_0}} \right.} \right) + \mathfrak{Q}
	\Big]\!,
\end{align}
where $\mathfrak{Q}\!\!\!=\!\!\!\sum\limits_{t = 2}^T \!{\log\! {q_\sigma }\!\left( {{{\mathbf{H}}_{t \!-\! 1}}\!\left| {{{\mathbf{H}}_t},{{\mathbf{H}}_0}} \right.} \right)} \!\! -\!\! \sum\limits_{t = 1}^T\! {\log \!{p_{\boldsymbol{\theta }}}\!\left( {{{\mathbf{H}}_{t \!-\! 1}}\left| {{{{\mathbf{H}}_t}},{{{{\mathbf{\hat H}}}^ \star }}} \right.} \right)}\! \!-\!\!\log\! {p_{\boldsymbol{\theta }}}\!\left(\! {{{\mathbf{H}}_T}} \!\right) $. We now have the following proposition to show that the GDM and the NM-GDM share the same objective, indicating that the proposed GDM can be further accelerated.

\textit{Proposition 1:} For all $\sigma  > 0$, there exist coefficients $\zeta  \in \mathbb{R}$, such that $\mathcal{L} = {{\mathcal{L}}_{\rm{Simp}} } + \zeta$. The NM-GDM degenerates to the GDM if and only if $\sigma _t^2 = {\beta _t}{{\left( {1 - {{\bar \alpha }_{t - 1}}} \right)} \mathord{\left/
		{\vphantom {{\left( {1 - {{\bar \alpha }_{t - 1}}} \right)} {\left( {1 - {{\bar \alpha }_t}} \right)}}} \right.
		\kern-\nulldelimiterspace} {\left( {1 - {{\bar \alpha }_t}} \right)}}$.

From Proposition 1, $\mathcal{L}$ is equivalent to ${{\mathcal{L}}_{\rm{Simp}} } $ to some extent, so the optimization direction of $\mathcal{L}$ is consistent with ${L_{\rm{Simp}} }$. In addition, the NM-GDM is no longer limited by the sampling interval caused by the Markovian chain assumption, and ${{\mathbf{H}}_{0}}$ can be generated by fewer sampling steps. In particular, let $\mathcal{T} = \left[ {T,T - 1,...,1} \right]$ be the original generated sequence with length $\dim \left( \mathcal{T} \right) = T$. Construct a time sub-sequence $\mathfrak{S} = \left[ {{\mathfrak{S}_S},{\mathfrak{S}_{S - 1}},...,{\mathfrak{S}_1}} \right]$ from $\mathcal{T}$ and a subset $\left\{ {{{\mathbf{H}}_{{\mathfrak{S}_1}}},...,{{\mathbf{H}}_{{\mathfrak{S}_S}}}} \right\}$ from ${{\mathbf{H}}_{1:T}}$, where $\mathfrak{S}$ is an increasing sub-sequence of $\left[ {1,...,T} \right]$, $\dim \left( \mathfrak{S} \right) = S$, and $S \leq T$. Then, we can define the sequential forward process over ${{\mathbf{H}}_{{\mathfrak{S}_1}}}$,..., ${{\mathbf{H}}_{{\mathfrak{S}_S}}}$ to match the marginal distribution \eqref{eq:marginal ddpm}, i.e., $q\left( {{{\mathbf{H}}_{{\mathfrak{S}_i}}}\left| {{{\mathbf{H}}_0}} \right.} \right) = \mathcal{N}\left( {{{\mathbf{H}}_{{\mathfrak{S}_i}}};\sqrt {1 - {\beta _{_{{\mathfrak{S}_i}}}}} {{\mathbf{H}}_0},{\beta _{_{{\mathfrak{S}_i}}}}{\mathbf{I}}} \right)$. Thus, in the generative process, the sampling time is much shorter than $T$.
\section{Network Architecture of the NM-GDM}
In this section, we present the network architecture of NM-GDM ${{{f }}_{\boldsymbol{\theta }}}\left(  \cdot  \right)$, which adopts the backbone of PixelCNN++ \cite{Ho11}. As shown in \figref{fig:model}, ${{f}}_{\boldsymbol{\theta }}\left(  \cdot  \right)$ mainly includes the Down Block and its corresponding inverse process, the Middle Block, and the Time Embedding Block. For clarity, we briefly introduce the components and working mechanism of each block.

\textit{1) Time Embedding Block:} Since our proposed model is related to time series $[1,..., t,...,T]$, to help the model better distinguish between different time constants, we augment information about the relative or absolute value of $t$ in the time series. To this end, we incorporate a time encoder to embed the time constant $t$. Specifically, akin to the Transformer sinusoidal position embedding \cite{vaswani2017attention}, we employ sine and cosine functions of varying frequencies to encode the time constant $t$, resulting in the time embedding vector ${{\bf{TE}}}_t$. Let ${{\rm{TE}}}^i_t \in {\mathbb{R}}$ be the $i$-th element of the time embedding vector for $t$, which is given by
\begin{align}\label{eq：49}
  {\rm{TE}}_t^{(i)} = \left\{ \begin{gathered}
  	\sin \left( {\frac{t}{{{{10000}^{{{2i} \mathord{\left/
  								{\vphantom {{2i} {{c_{{\rm{time}}}}}}} \right.
  								\kern-\nulldelimiterspace} {{c_{{\rm{time}}}}}}}}}}} \right),\text{ if}\quad i = 2k \hfill \\
  	\cos \left( {\frac{t}{{{{10000}^{{{2i} \mathord{\left/
  								{\vphantom {{2i} {{c_{{\rm{time}}}}}}} \right.
  								\kern-\nulldelimiterspace} {{c_{{\rm{time}}}}}}}}}}} \right),\text{ if}\quad i = 2k + 1 \hfill \\ 
  \end{gathered}  \right.,
\end{align}
where $i = 0,1,2,...,{c_{{\rm{time}}}}/2-1$, $c_{{\rm{time}}}$ is the dimension of the time embedding vector. Combined with \eqref{eq：49}, the embedding vector of the time constant $t$ is given by
\begin{align}\label{eq：50}
	{{\bf{TE}}_t} &= \Big[\sin \big( {{w_0}t} \big),\cos \big( {{w_0}t} \big),\sin \big( {{w_1}t} \big),\cos \big( {{w_1}t} \big),\nonumber \\
	&\qquad\qquad...,\sin \big( {{w_{\tfrac{{{c_{{\rm{time}}}}}}{2} - 1}}t} \big),\cos \big( {{w_{\tfrac{{{c_{{\rm{time}}}}}}{2} - 1}}t} \big)\Big],
\end{align}
where ${w_i} = {1 \mathord{\left/
		{\vphantom {1 {{{10000}^{{{2i} \mathord{\left/
									{\vphantom {{2i} {{c_{{\text{time}}}}}}} \right.
									\kern-\nulldelimiterspace} {{c_{{\rm{time}}}}}}}}}}} \right.
		\kern-\nulldelimiterspace} {{{10000}^{{{2i} \mathord{\left/
						{\vphantom {{2i} {{c_{{\rm{time}}}}}}} \right.
						\kern-\nulldelimiterspace} {{c_{{\text{time}}}}}}}}}}$. Note that the embedding vectors ${\bf{TE}}_{t + \Delta t}$ for different time constants $t + \Delta t$ can be obtained by a linear variation of ${{\bf{TE}}_t}$, i.e.,
\begin{align}\label{eq：51}
	{\bf{TE}}_{t + \Delta t}^T = \left[ {\begin{array}{*{20}{c}}
			{\sin \Big( {{w_0}\big( {t + \Delta t} \big)} \Big)} \\ 
			{\cos \Big( {{w_0}\big( {t + \Delta t} \big)} \Big)} \\ 
			...\\
			{\sin \Big( {{w_{\tfrac{{{c_{{\rm{time}}}}}}{2} - 1}}\big( {t + \Delta t} \big)} \Big)} \\
			{\cos \Big( {{w_{\tfrac{{{c_{{\rm{time}}}}}}{2} - 1}}\big( {t + \Delta t} \big)} \Big)} 
	\end{array}} \right] = {{\bf{D}}_{\Delta t}} \cdot {\bf{TE}}_t^T,
\end{align}
where ${\bf{D}}_{\Delta t}$ is a transform matrix defined as \eqref{eq:52}, shown at the bottom of the next page.
\begin{figure*}[!b]
	\hrule
	\begin{align}\label{eq:52}
		{{\mathbf{D}}_{\Delta t}} = \left( {\begin{array}{*{20}{c}}
				{\left[ {\begin{array}{*{20}{c}}
							{\cos \left( {{w_0}\Delta t} \right)}&{\sin \left( {{w_0}\Delta t} \right)} \\ 
							{ - \sin \left( {{w_0}\Delta t} \right)}&{\cos \left( {{w_0}\Delta t} \right)} 
					\end{array}} \right]}&{...}&0 \\ 
				{...}&{...}&{...} \\ 
				0&{....}&{\left[ {\begin{array}{*{20}{c}}
							{\cos \left( {{w_{\tfrac{{{c_{{\rm{time}}}}}}{2} - 1}}\Delta t} \right)}&{\sin \left( {{w_{\tfrac{{{c_{{\rm{time}}}}}}{2} - 1}}\Delta t} \right)} \\ 
							{ - \sin \left( {{w_{\tfrac{{{c_{{\rm{time}}}}}}{2} - 1}}\Delta t} \right)}&{\cos \left( {{w_{\tfrac{{{c_{{\rm{time}}}}}}{2} - 1}}\Delta t} \right)} 
					\end{array}} \right]} 
		\end{array}} \right)
	\end{align}
\end{figure*}
Combining \eqref{eq：50} and \eqref{eq：51}, we can encode any time constant $t$ into the time embedding vector ${\bf{TE}}_t$, and then further embedding is completed by two linear layers and one activation layer, i.e., ${\mathbf{T}}{{{\mathbf{E'}}}_t} = {f_{{\rm{Lin}}}}\left( {{f_{{\rm{Sig}}}}\left( {{f_{{\rm{Lin}}}}\left( {{\mathbf{T}}{{\mathbf{E}}_t}} \right)} \right)} \right)$,
where ${f_{{\rm{Lin}}}}\left(  \cdot  \right)$ and ${f_{{\rm{Sig}}}}\left(  \cdot  \right)$ represent the linear layer and sigmoid activation function layer, respectively.

\begin{figure*}[!t]
	\centering
	\includegraphics[scale=0.176]{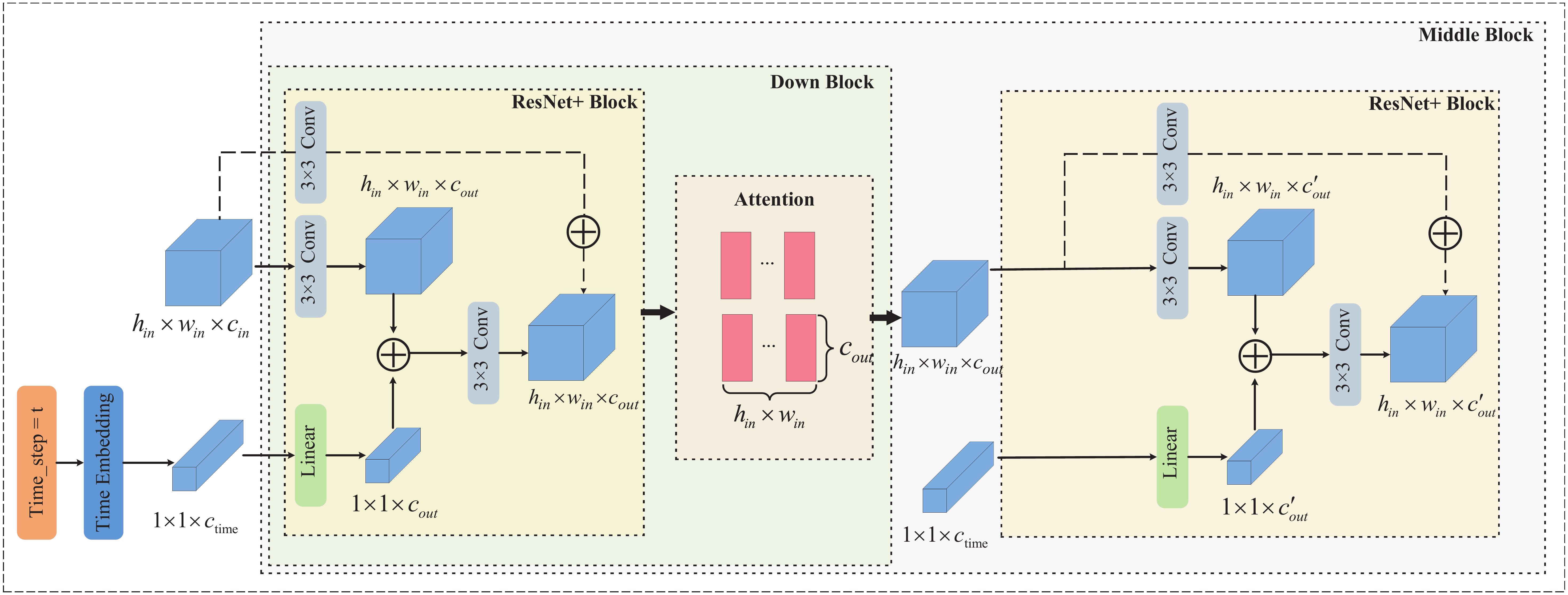}
	\captionsetup{font=footnotesize}
	\caption{Key blocks of the proposed NM-GDM. The diagram mainly shows the essential modules: ResNet+ block, Down (or Up), and Middle block.}
	\label{fig:3}
\end{figure*}

\textit{2) ResNet+ Block:} To improve the network's representation capability, we establish deeper feature extraction networks to learn the higher-level semantic information of the XL-MIMO channels. Deep neural networks usually encounter model degradation due to issues such as vanishing gradients and exploding gradients. Hence, we incorporate residual connections as a strategy to alleviate the above problems. As shown in \figref{fig:3}, let ${\bf{X}} \in {\mathbb{R}^{{h_{{\rm{in}}}} \times {w_{{\rm{in}}}} \times {c_{{\rm{in}}}}}}$, ${\bf{TE}}_t$ be the feature map and time embedding vector inputs, respectively. The output ${\bf{X}}'\in {\mathbb{R}^{{h_{{\rm{in}}}} \times {w_{{\rm{in}}}} \times {c_{{\rm{out}}}}}}$ of ResNet+ block is given by
\begin{align}\label{eq：}
	{\bf{X}}' = \left\{ \begin{gathered}
		\boldsymbol{\wp}  + {{\tilde f}_{{\rm{Conv}}}}\left( {\bf{X}} \right),\text{ if }{c_{{\rm{out}}}} \ne {c_{{\rm{in}}}} \hfill \\
		\boldsymbol{\wp}  + {\bf{X}},\text{ else} \hfill \\ 
	\end{gathered}  \right.,
\end{align}
where ${\boldsymbol{\wp}}  \in {\mathbb{R}^{{h_{{\rm{in}}}} \times {w_{{\rm{in}}}} \times {c_{{\rm{out}}}}}}$ is an intermediate variable in the feature extraction process. ${\boldsymbol{\wp}}  = {f_{{\rm{Conv}}}}\left( {{f_{{\rm{Dro}}}}\left( {{f_{{\rm{Sig}}}}\left( {{f_{{\rm{Gn}}}}\left( {g({\bf{X}}) + {f_{{\rm{Sig}}}}\left( {{f_{{\rm{Lin}}}}\left( {{\mathbf{T}}{{{\mathbf{E'}}}_t}} \right)} \right)} \right)} \right)} \right)} \right)$,
where $g({\bf{X}}) = {f_{{\rm{Conv}}}}\left( {{f_{{\rm{Sig}}}}\left( {{f_{{\rm{Gn}}}}\left( {\bf{X}} \right)} \right)} \right)$, ${f_{{\rm{Conv}}}}\left(  \cdot  \right)$, ${{\tilde f}_{{\rm{Conv}}}}\left(  \cdot  \right)$ are $3\times3$ and $1\times1$ convolution operations, respectively. ${f_{{\rm{Dro}}}}\left(  \cdot  \right)$, ${f_{{\rm{Gn}}}}\left(  \cdot  \right)$  are the dropout layer and group normalization layer functions, respectively. Note that the dropout layer function can enhance the diversity of model learning and reduce the complex co-adaptation relationship between neurons.

\textit{3) Attention Block:} To encode the global structural information of the XL-MIMO channels, we establish the attention block to enhance the interaction between global and local features. Specifically, to implement the self-attention mechanism, a normalized attention matrix is introduced to represent varying degrees of attention to the input. More significant input components are assigned higher weights. The ultimate output is computed by weighting the input based on the attention weights specified in the attention matrix. Specifically, given the input $\mathbf{Z}\! =\! {[{\mathbf{z}_1},...,{\mathbf{z}_m}]^T} \in {\mathbb{R}^{{d_m} \times {d_n}}}$, ${{d_m}}$ denotes the number of image patches and ${{d_n}}$ is the feature dimension of each patch, three different linear transformations are applied to ${\mathbf{z}_j}$ \cite{vaswani2017attention}:
\begin{subequations}\label{eq:spdb}
	\begin{align}
		{\mathbf{k}_j} &= {\mathbf{z}_j}{\mathbf{W}^k},\quad j = 1,...,{d_n},\label{eq:dbso}\\
		{\mathbf{q}_j} &= {\mathbf{z}_j}{\mathbf{W}^q},\quad j = 1,...,{d_n},\label{eq:dbsc}\\
		{\mathbf{v}_j} &= {\mathbf{z}_j}{\mathbf{W}^v},\quad j = 1,...,{d_n},\label{eq:dbss} 
	\end{align}
\end{subequations}
where ${\mathbf{k}_j} \in {\mathbb{R}^{1 \times {d_k}}}$, ${\mathbf{q}_j} \in {\mathbb{R}^{1 \times {d_q}}}$, and ${\mathbf{v}_j} \in {\mathbb{R}^{1 \times {d_n}}}$ are the key, query, and value vector, respectively. ${\mathbf{W}^k} \in {\mathbb{R}^{{d_n} \times {d_k}}}$, ${\mathbf{W}^q} \in {\mathbb{R}^{{d_n} \times {d_q}}}$, and ${\mathbf{W}^v} \in {\mathbb{R}^{{d_n}\times {d_n}}}$ represent the respective trainable transformation matrices, with ${d_k} = {d_q}$. In detail, the weight allocation function is determined by ${\mathbf{k}_j}$ and ${\mathbf{q}_{j'}}$. A higher correlation ${\mathbf{q}_{j'}}\mathbf{k}_j^T$ implies that the features of the $j$-th input patch ${\mathbf{z}_j}$ hold greater importance for the ${j'}$-th output patch. Generally, this correlation can be adaptively adjusted based on the input $\mathbf{Z}$ and the matrices ${\mathbf{W}^k}$ and ${\mathbf{W}^q}$. For clarity, the matrix forms of \eqref{eq:dbso}-\eqref{eq:dbss} are presented as ${\mathbf{K}} = {\mathbf{Z}}{\mathbf{W}^k}$, ${\mathbf{Q}} = {\mathbf{Z}}{\mathbf{W}^q}$, ${\mathbf{V}} = {\mathbf{Z}}{\mathbf{W}^v}$,
where $\mathbf{K} = {[{\mathbf{k}_1},...,{\mathbf{k}_m}]^T} \in {\mathbb{R}^{{d_m} \times {d_k}}}$, $\mathbf{Q} = {[{\mathbf{q}_1},...,{\mathbf{q}_m}]^T} \in {\mathbb{R}^{{d_m} \times {d_q}}}$ and $\mathbf{V} = {[{\mathbf{v}_1},...,{\mathbf{v}_m}]^T} \in {\mathbb{R}^{{d_m} \times {d_n}}}$.

Leveraging $\mathbf{K}$ and $\mathbf{Q}$, we can obtain the attention matrix $\mathbf{AM}  \in {\mathbb{R}^{{d_m} \times {d_m}}}$, which is denoted as 
\begin{align}\label{eq:cchmd}
	\mathbf{AM} = \rm{Softmax} \left( {\frac{{\mathbf{Q}{\mathbf{K}^\textit{T}}}}{{\mathit{\sqrt \eta} }}} \right),
\end{align}
where $\rm{Softmax} \left( \mathbf{Z}\right)  = {\exp \left( \mathit{{z_j}}\right) }/{{\sum {\exp \left( \mathit{{z_j}}\right) } }}$, and $\sqrt \eta >0$. Each column of the attention matrix is a vector of attention scores, i.e., each score is a probability, where all scores are non-negative and sum up to 1. Note that when the key vector $\mathbf{K}[j,:]$ and the query $\mathbf{Q}[j',:]$ has a better match, the corresponding attention score $\mathbf{AM}[j',j]$ is higher. Thus, the output of the attention mechanism corresponding to the $r$-th component can be represented by the weighted sum of all inputs, i.e., ${\mathbf{z}'_r} = \sum\nolimits_j {\mathbf{AM} [r,j]{\mathbf{v}_j}}  = \mathbf{AM} [r,:] \cdot \mathbf{V}$, where ${\mathbf{z}'_r} \in {\mathbb{R}^{1 \times {d_n}}}$ represents the $r$-th output, which is computed by adaptively focusing on the inputs based on $\mathbf{AM}[r,j]$. When $\mathbf{AM}[r,j]$ is higher, the associated value vector ${\mathbf{v}_j}$ will have a more significant impact on the $r$-th output patch. Finally, the output ${\bf{Z}}'$ of the attention block is given by 
\begin{align}\label{eq:cmd}
	{\bf{Z}}' &=\mathbf{AM}[:,:]\cdot \mathbf{V}\nonumber \\
	&=\rm{Softmax}\mathit{\left( {\frac{{\mathbf{Z}{\mathbf{W}^q}{\mathbf{W}^{{k^T}}}{\mathbf{Z}^T}}}{{\sqrt \eta }}} \right)\mathbf{Z}{\mathbf{W}^v}},
\end{align}
where $\mathbf{O} = {[{\mathbf{o}_{1,...,}}{\mathbf{o}_m}]^T} \in {\mathbb{R}^{{d_m} \times {d_n}}}$.
\newcolumntype{L}{>{\hspace*{-\tabcolsep}}l}
\newcolumntype{R}{c<{\hspace*{-\tabcolsep}}}
\definecolor{lightblue}{rgb}{0.93,0.95,1.0}
\begin{table}[!b]
	\captionsetup{font=footnotesize}
	\caption{Wireless System and Model Setup Parameters}\label{ta:sys}
	\centering
	\setlength{\tabcolsep}{16mm}
	\ra{1.9}
	\scriptsize
	\scalebox{0.8}{\begin{tabular}{LR}
			\toprule
			Parameter &  Value\\
			\midrule
			\rowcolor{lightblue}
			Number of BS antennas & ${N}=256$  \\
			Number of Users & ${M}=16$    \\
			\rowcolor{lightblue}
			Number of RF chains & ${N_{\rm{RF}}}=16$\\
			Carrier frequency & ${f_c}=28$ GHz\\
			\rowcolor{lightblue}
			Bandwidth & $B=100$ MHz\\
			Number of subcarriers & $K=256$ \\
			\rowcolor{lightblue}
			Number of channel paths for per user& $L=6$\\	
			Number of feature channels in the latent space & ${c_1}=\left\{ {16,32,64} \right\}$  \\
			Ratio of $c_1$, $c_2$, $c_3$, $c_4$, & ${\eta _c} = 1:2:2:2$    \\
			\rowcolor{lightblue}
			Number of ResNet+ blocks & $N_{\rm{RB}}=3$\\
			Dropout rate & ${d_r}=0.1$ \\
			\rowcolor{lightblue}
			EMA rate& ${{e_r}}=0.9999$ \\
			Learning rate& ${\alpha }=0.0001$ \\						
			\bottomrule
		\end{tabular}
	}
\end{table}

Note that the proposed Down (or Up) block is composed of $N_{{\rm{RB}}}$ ResNet+ blocks and the corresponding attention blocks, and the middle block is composed of two ResNet+ blocks, and an attention block, as summarized in \figref{fig:3}.
\section{Numerical Experiment}
In this section, a series of experiments are carried out to evaluate the performance of our approach. Initially, the implementation details are introduced. Then, we analyze the convergence and complexity of the proposed model across various hyperparameter settings. Finally, we compare the proposed approach and baselines under typical system configurations.
\subsection{Implementation Details}
We consider an uplink XL-MIMO system where the BS is equipped with a ULA of $N=256$ antennas to serve $M=16$ single-antenna users. The BS adopts a hybrid digital-analog architecture with a limited number of $N_{{\rm{RF}}}=16$ RF chains, operating at a carrier frequency of $f_c=28$ GHz and a bandwidth of $B=100$ MHz. Additionally, the elements within ${\mathbf{C}}$ in \eqref{eq:9} are independent and randomly selected from the set $1/{\sqrt{N}}\left\{ { - 1,1} \right\}$ with equal probability \cite{7790909,9693928}. The SNR is defined as $1/{\sigma ^2}$ \cite{9693928}. Based on the system configuration parameters presented in \tabref{ta:sys}, we generate the corresponding channel data pairs $\left\{ {{{{{\mathbf{\hat H}}}^ \star }}_{i},{{\mathbf{H}}}_{i}} \right\}_{i = 1}^{70000}$, where the NF channel ${{\mathbf{H}}}$ is generated following the method in \cite{9693928} to ensure a fair comparison of performance in subsequent experiments, with ${{{{\mathbf{\hat H}}}^ \star }}$ obtained using the SOMP algorithm. The training, validation, and testing sets are divided in a $5:1:1$ ratio. Furthermore, we scale the data to the [0,1] range for training.

On the hardware level, the proposed model is trained on 2 Nvidia RTX-4090 GPU with 24 GB of memory and tested on 1 Nvidia RTX-4090 GPU with 24 GB of memory. On the algorithm level, the Adam optimizer is adopted, with the initial learning rate set to $\alpha=0.0001$. The exponential moving average algorithm is adopted, with the decay factor set to 0.9999. Besides, we introduce the dropout algorithm \cite{Ho11}. The proposed model is trained by minimizing the objective function \eqref{eq：47} with $\gamma=1$, and the count of time steps, $T$, in the forward process is set to 1000. The variance of ${q_\sigma }\left( {{{\mathbf{H}}_{t - 1}}\left| {{{\mathbf{H}}_t},{{\mathbf{H}}_0}} \right.} \right)$ is set to 0 to accelerate the sampling, and the count of sampling steps, $S$, is set to 100. The level of the diffusion noise adheres to a linear variance schedule that is evenly spaced, starting from $\beta _1=10^{-4}$ to $\beta _T= 0.02$. The batch size is set to 512, and the model parameters are optimized over 500,000 iterations. Additional specifics on model parameters are detailed in \tabref{ta:sys}.

\subsection{Convergence and Complexity Analysis}
\begin{figure*}[!b]
	\centering
	\subfloat[Model performance versus $\alpha$.]{\includegraphics[width=0.25\textwidth]{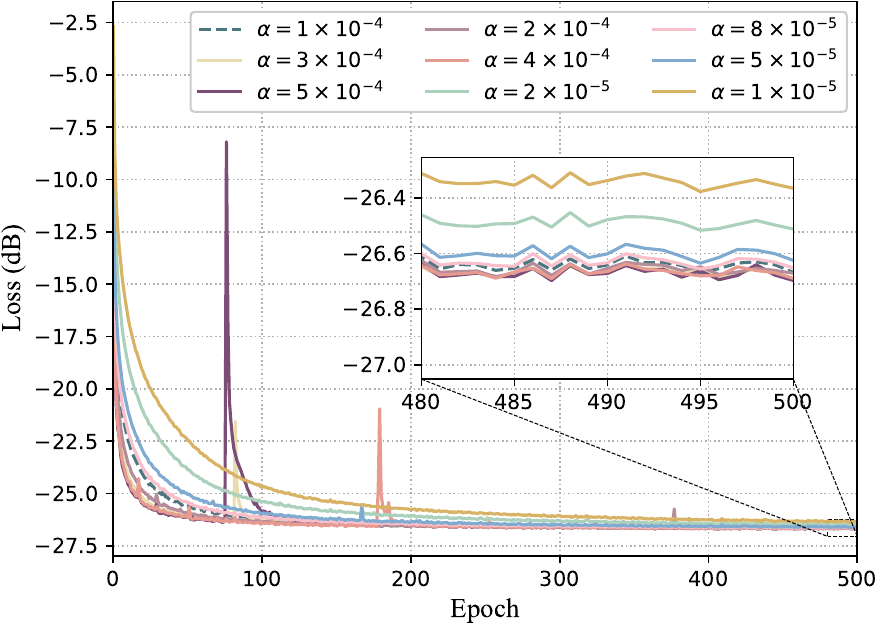}\label{fig:lr}}
	\hfill
	\subfloat[Model performance versus $d_r$.]{\includegraphics[width=0.25\textwidth]{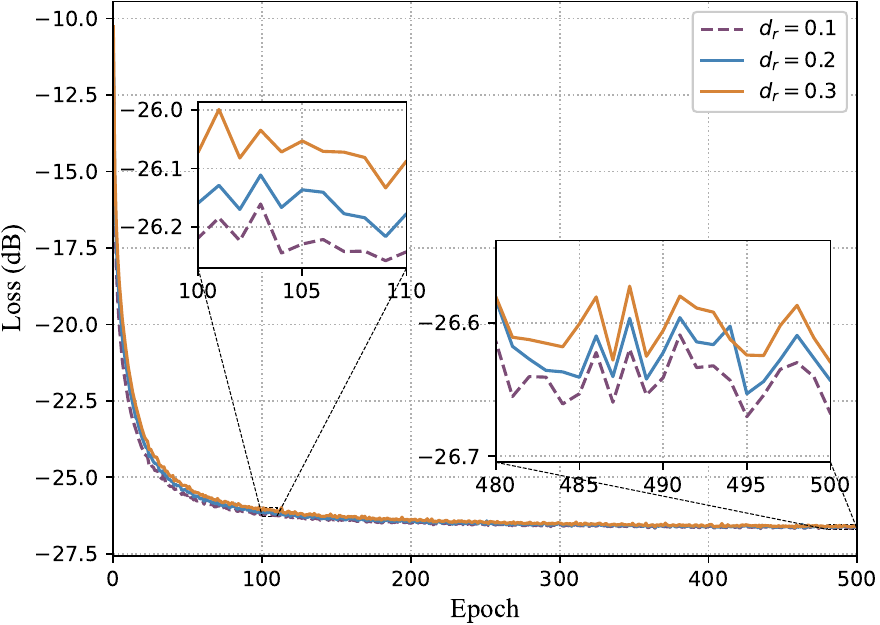}\label{fig:dr}}
	\hfill
	\subfloat[Model performance versus $c_1$.]{\includegraphics[width=0.25\textwidth]{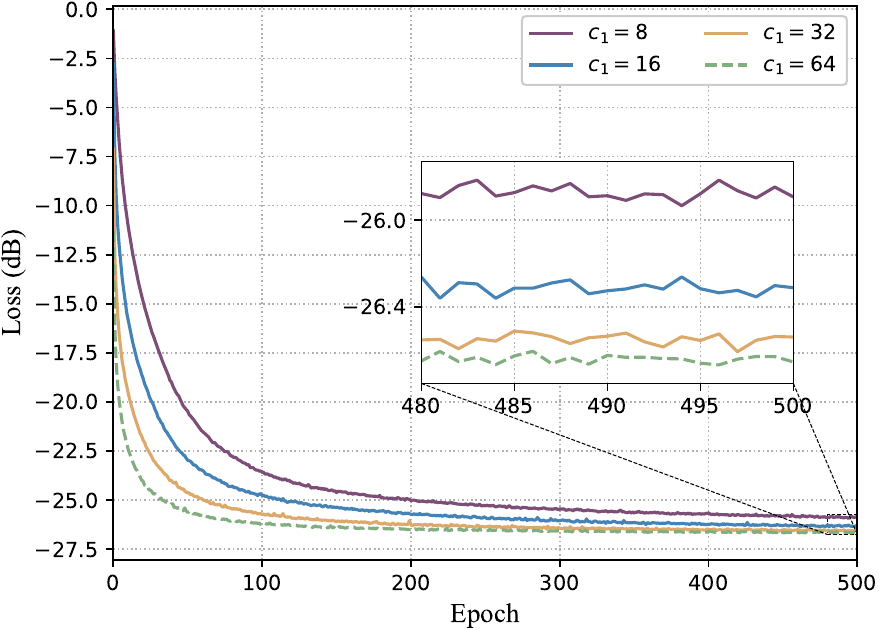}\label{fig:c_1}}
	\hfill
	\subfloat[Model performance versus ${\eta _c}$.]{\includegraphics[width=0.25\textwidth]{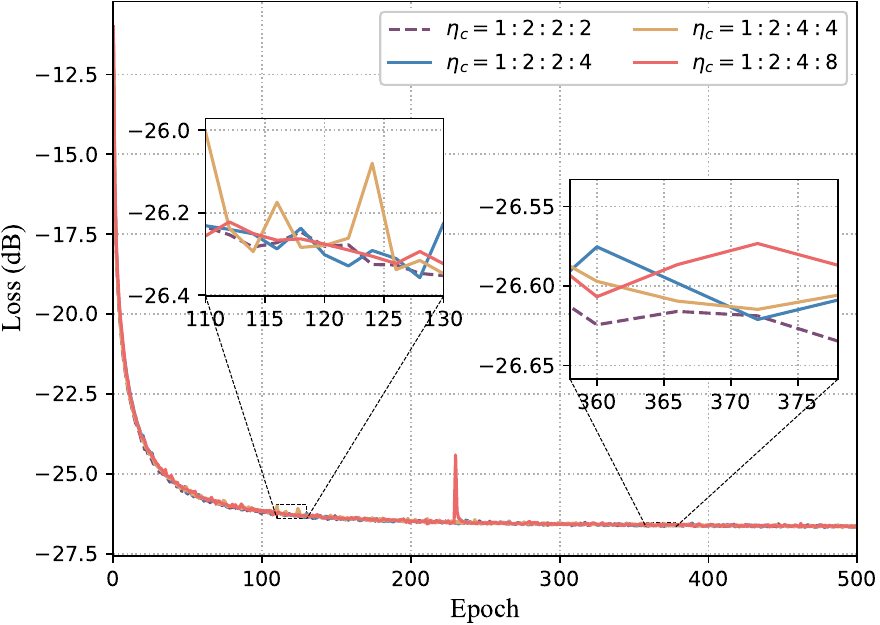}\label{fig:ch_mul}}
	\captionsetup{font=footnotesize}
	\caption{Loss analysis of NM-GDM for different hyperparameter settings.}
	\label{fig:loss}
\end{figure*}

\subfigref{fig:loss}{fig:lr} compares the training performance of our models across various values of the learning rate $\alpha$. When $\alpha$ is set to $5 \times 10^{-4}$, the objective function curve of the model exhibits multiple spikes during the training process, and the oscillation amplitude of the curve is significant. These similar curve variations also occur when $\alpha$ is set to $4 \times {10^{ - 4}}, 3 \times {10^{ - 4}}, 2 \times {10^{ - 4}}$, respectively. It is found that when $\alpha$ is set exceedingly large, the model is difficult to converge as it is prone to getting trapped in local optima. In contrast, during the later stages of training, the loss associated with $\alpha=1 \times {10^{ - 5}}$ is the highest among all $ \alpha$ settings, which implies that when $\alpha$ is set excessively small, the model converges very slowly, necessitating more training epochs to acquire the optimal solution. Besides, an excessively small $\alpha$ may cause the model to misinterpret a local minima for the global optimum and become trapped. Furthermore, when $\alpha$ is set to $1 \times {10^{ - 4}}$, the corresponding loss curve shows no apparent spikes, demonstrating impressive convergence and estimation accuracy. Hence, $\alpha=1 \times {10^{ - 4}}$ is selected as the default for the subsequent experiments.

\subfigref{fig:loss}{fig:dr} compares the effect of different $d_r$ on the performance of the proposed model. According to the general experience, the proposed model is trained with $d_r=0.1$, $0.2$, and $0.3$, respectively. It can be found that when $d_r$ is set large, i.e., $d_r=0.3$, the convergence speed of the proposed model is slow, and its accuracy is lower than that of the model with $d_r=0.1$ and $0.2$. We believe that when the value of $d_r$ is exceedingly large, the model may lose some important features during the learning process. \subfigref{fig:loss}{fig:c_1} compares the influence of $c_1$ on the proposed model. When $c_1$ is increased to 64, the proposed model achieves better convergence speed and accuracy. \subfigref{fig:loss}{fig:ch_mul} demonstrates the impact of different $\eta_c$ on the proposed model. Doubling the feature-channel dimension of the feature maps in the intermediate layer, such as ${\eta _c} = 1:2:4:8$, is not necessarily the best option. Based on observation and experience, ${\eta _c} = 1:2:2:2$ is the optimal choice for our model.
\begin{figure}[!t]
	\centering
	\subfloat[$N_\mathrm{RB}=3$.]{\includegraphics[width=0.43\textwidth]{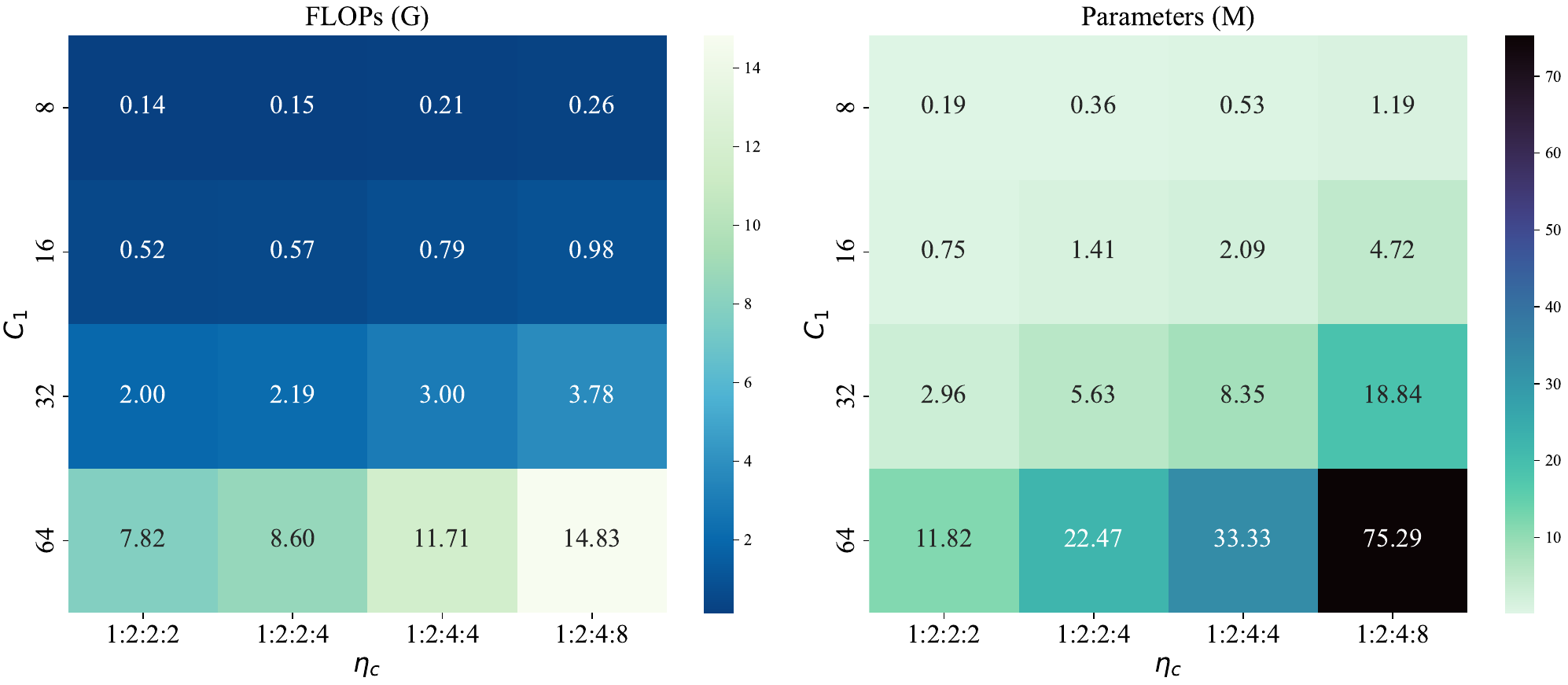}\label{fig:nrb}}\\
	\subfloat[$c_1=16$.]{\includegraphics[width=0.43\textwidth]{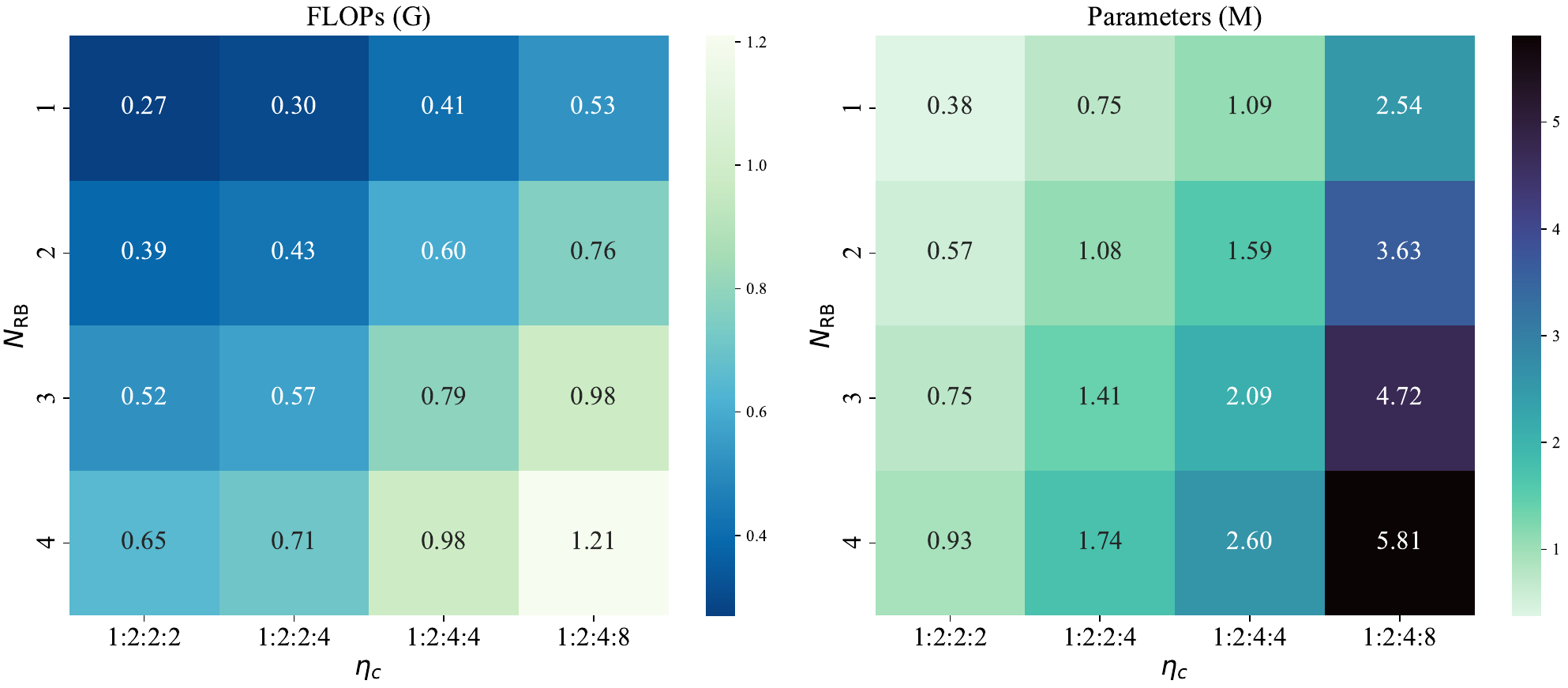}\label{fig:c1}}\\
	\subfloat[${\eta _c=1:2:2:4}$.]{\includegraphics[width=0.43\textwidth]{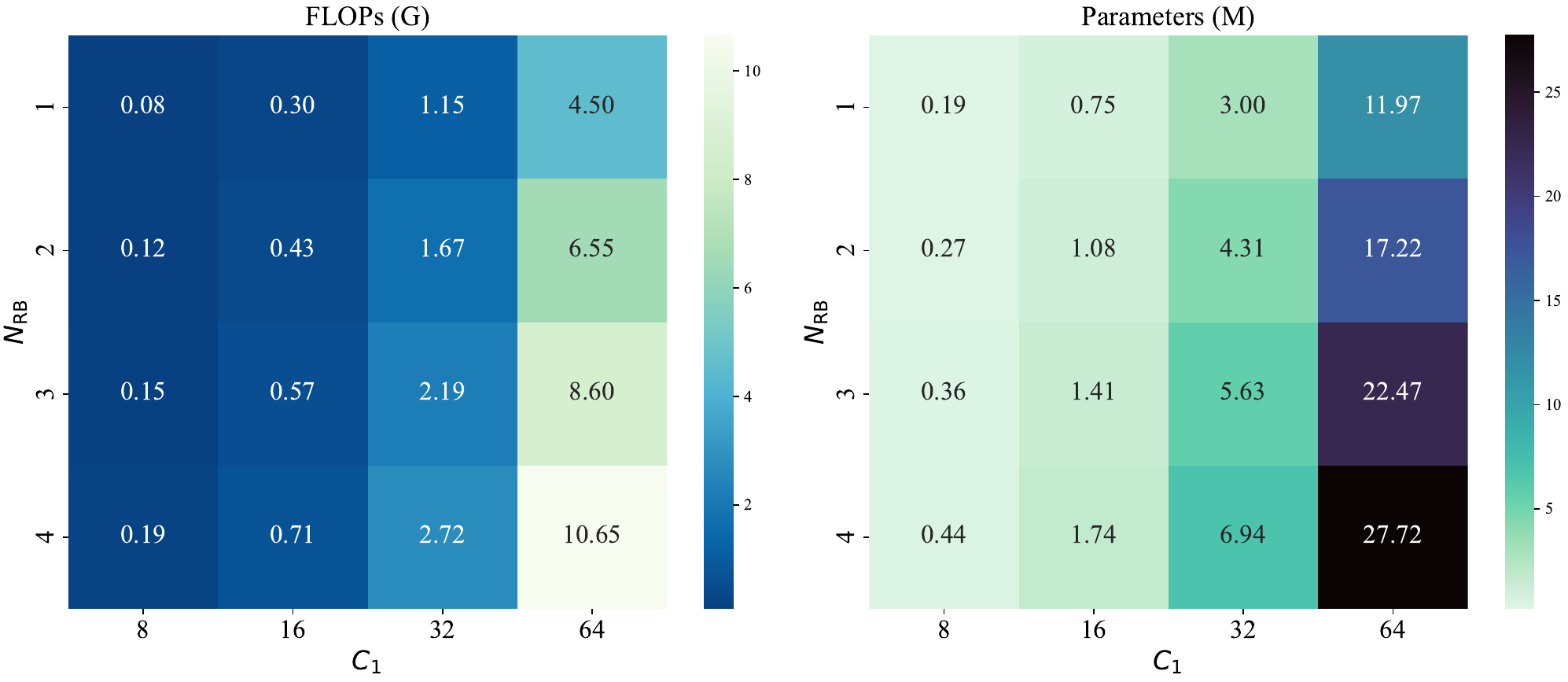}\label{fig:etac}}
	\captionsetup{font=footnotesize}
	\caption{Analysis of the NM-GDM's computational complexity (GFLOPs) and storage complexity (model parameters)}
	\label{fig:flops}
\end{figure}
\begin{figure}[!b]
	\centering
	\includegraphics[width=0.41\textwidth]{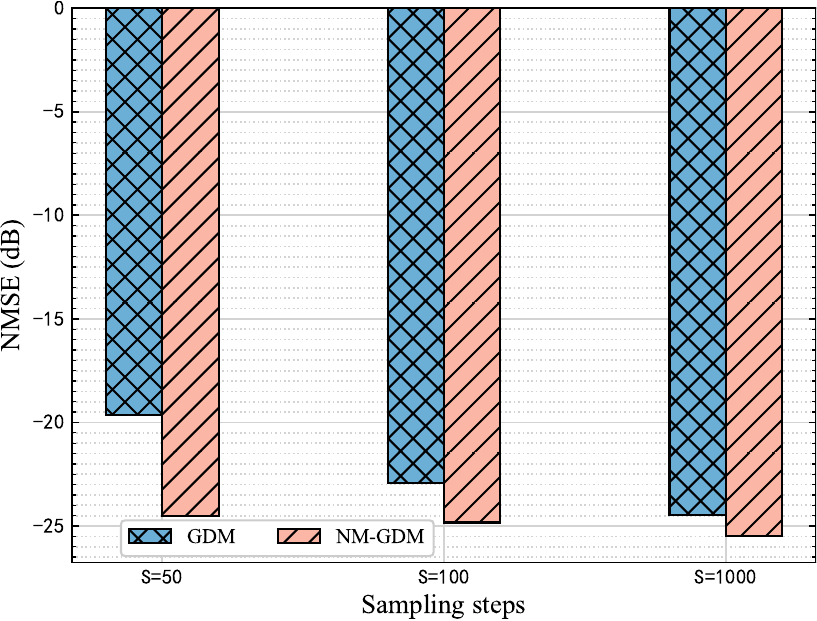}
	\captionsetup{font=footnotesize}
	\caption{Comparison of NMSE performance of the proposed GDM and NM-GDM over different sampling steps.}
	\label{fig:sample}
\end{figure}
\begin{figure}[!b]
	\centering
	\subfloat[Users are located at (15 m, 25 m).]{\includegraphics[width=0.41\textwidth]{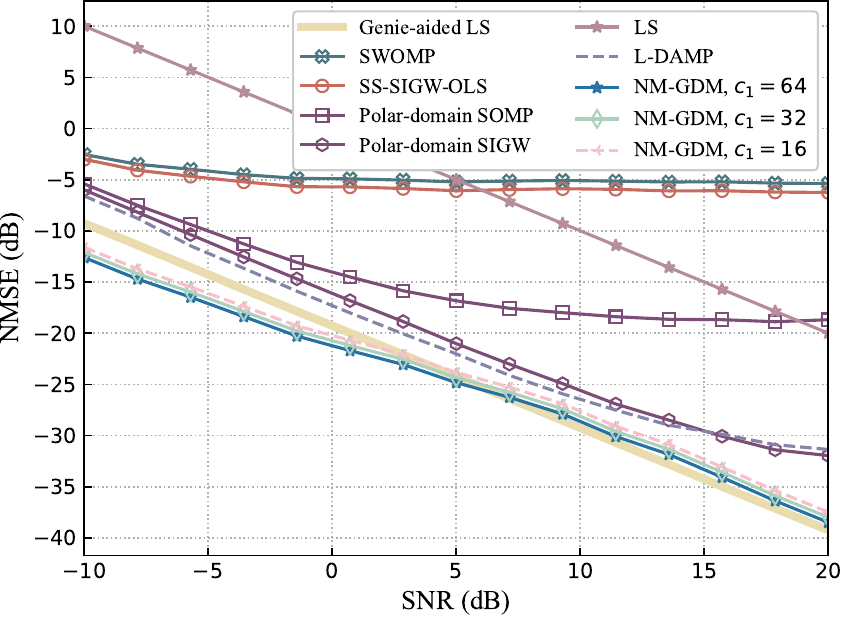}\label{fig:snr_r=18}}\\
	\subfloat[Users are located at (350 m, 400 m).]{\includegraphics[width=0.41\textwidth]{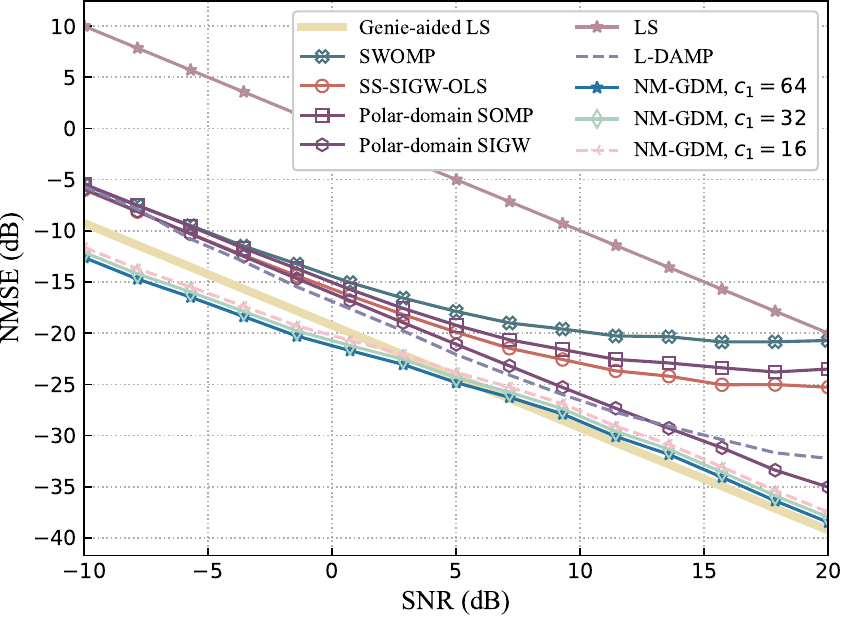}\label{fig:snr_r=398}}
	\captionsetup{font=footnotesize}
	\caption{Comparison of the NMSE performance of the NM-GDM and the baselines at different SNRs for NF (a) and FF (b) regions when $Q=32$.}
	\label{fig:snr}
\end{figure}

Given that the proposed method adopts a two-stage framework, we provide a corresponding complexity analysis for each stage individually. The computational complexity of the initial CE stage can be directly obtained as the $\mathcal{O}\left( { LQ{N_{\mathrm{RF}}}SK} \right)$ by referring to the OMP algorithm \cite{9693928}. For the enhanced CE, we employ a denoising neural network model based on a UNet variant, as illustrated in \figref{fig:model}. To ensure a fairer evaluation of this stage, we adopt two widely recognized complexity metrics: model parameters for storage complexity and floating-point operations (FLOPs) for computational complexity. The storage and computational complexities of the proposed model are primarily determined by the architectural parameters, including the channel number multiplier ${\eta _c} = c_1:c_2:c_3:c_4$, the number $c_1$ of base channels, and the number $N_{\rm{RB}}$ of ResNet+ blocks. As illustrated in \figref{fig:flops}, we present the impact of varying ${\eta_c}$, $c_1$, and $N_{\rm{RB}}$ settings on the model’s parameter count (in millions) and FLOPs (in Giga). Regarding computational complexity, $c_1$ has a more pronounced impact on the model's FLOPs compared to $N_{\rm{RB}}$ and ${\eta _c}$. A similar impact is observed in the model's parameter count as well. It is evident that increasing $c_1$ leads to higher computational and storage complexities, which, as shown in previous experiments, also enhances the model's estimation performance. After considering the trade-off between performance and complexity, we select $N_{\rm{RB}}=3$, $c_1=64$, and ${\eta _c} = 1:2:2:2$ for the proposed model, resulting in 7.82 GFLOPs and 11.82 million parameters. Note that, for the proposed model, the input size does not affect the number of model parameters due to its fully convolutional nature, which provides a distinct advantage for scaling to larger MIMO dimensions.

Additionally, since the NM-GDM we employ is an iterative refinement approach, we also introduce the sampling steps as an additional metric for evaluating the model’s complexity. \figref{fig:sample} illustrates the NMSE performance of the proposed GDM and NM-GDM with respect to the sampling steps, where the SNR is 5 dB, the users are located at (15 m, 25 m), and the pilot length $Q$ is 32. $c_1$ is set to 64, and other model parameters are set according to \tabref{ta:sys}. In \figref{fig:sample}, a downward trend is observed as the count of sampling steps grows. It appears that when the value of $S$ is small, the corresponding CE performance of the GDM is relatively poor. This poor performance is because of the limitation of the Markov chain-like structure, which prohibits sampling with a large interval. Additionally, the performance corresponding to the NM-GDM with 100 sampling steps is nearly equivalent to that of the GDM with 1000 sampling steps, which validates our theoretical analysis and reasoning. In terms of sampling speed, the proposed NM-GDM is about 10 times faster than the GDM, and can be about 20 times faster when the requirement for CE accuracy is not highly stringent.
\begin{figure}[!b]
	\centering
	\subfloat[Users are located at (15 m, 25 m).]{\includegraphics[width=0.41\textwidth]{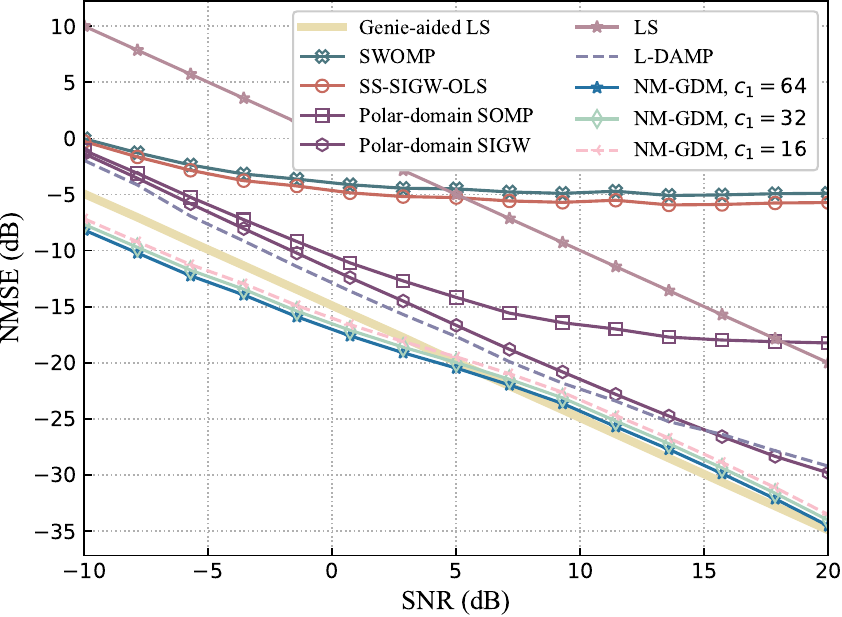}\label{figure:}}\\
	\subfloat[Users are located at (350 m, 400 m).]{\includegraphics[width=0.41\textwidth]{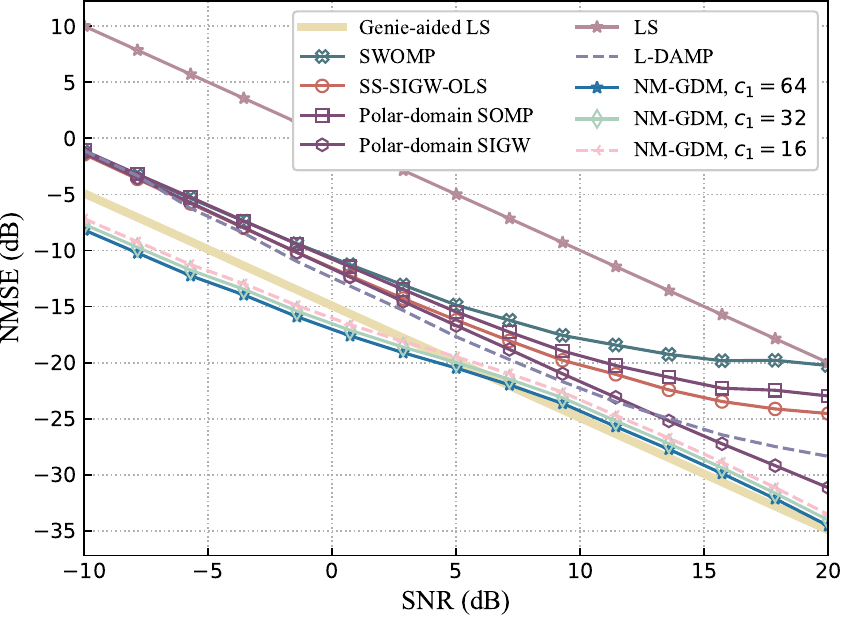}\label{figure:}}
	\captionsetup{font=footnotesize}
	\caption{Comparison of the NMSE performance of the NM-GDM and the baselines at different SNRs for NF (a) and FF (b) regions when $Q=8$.}
	\label{fig:SNR_PILOT}
\end{figure}

\subsection{Performance Comparison and Ablation Analysis}
In this subsection, we compare the proposed NM-GDM with the state-of-art algorithms, including the CNN-based L-DAMP \cite{he2018deep}, the polar-domain SOMP \cite{9693928} and SIGW algorithms \cite{9693928}, the angular-domain SW-OMP \cite{8306126} and SS-SIGW-OLS \cite{9246294} algorithms, and the LS algorithm. Additionally, the Genie-aided LS algorithm is adopted as a bound on the performance, which assumes that angular and distance information is available. Note that unless otherwise specified, the default settings are $c_1 = 64$, ${\eta_c} = 1:2:2:2$, and $\alpha = 1 \times 10^{-4}$ for the proposed NM-GDM. Without loss of generality, the CE performance of all approaches is evaluated by the NMSE metric, denoted as $\rm{NMSE} \!\!=\!\! \mathbb{E}\left( {{{\left\| {{{\mathbf{H}}} - {{{\mathbf{\hat H}}}^\prime }} \right\|_{\text{2}}^{\text{2}}} \mathord{\left/	
			{\vphantom {{\left\| {{{\mathbf{H}}} - {{{\mathbf{\hat H}}}^\prime }} \right\|_{\text{2}}^{\text{2}}} {\left\| {{{\mathbf{H}}}} \right\|_{\text{2}}^{\text{2}}}}} \right.
			\kern-\nulldelimiterspace} {\left\| {{{\mathbf{H}}}} \right\|_{\text{2}}^{\text{2}}}}} \right)$. Note that the first stage of our proposed two-phase approach, referred to as NM-GDM, utilizes the polar-domain SOMP algorithm \cite{9693928}. Consequently, the proposed NM-GDM and the polar-domain SOMP algorithm naturally form an ablation study.

\begin{figure*}[!t]
	\centering
	\subfloat[Users are located at (15 m, 25 m).]{\includegraphics[width=0.33\textwidth]{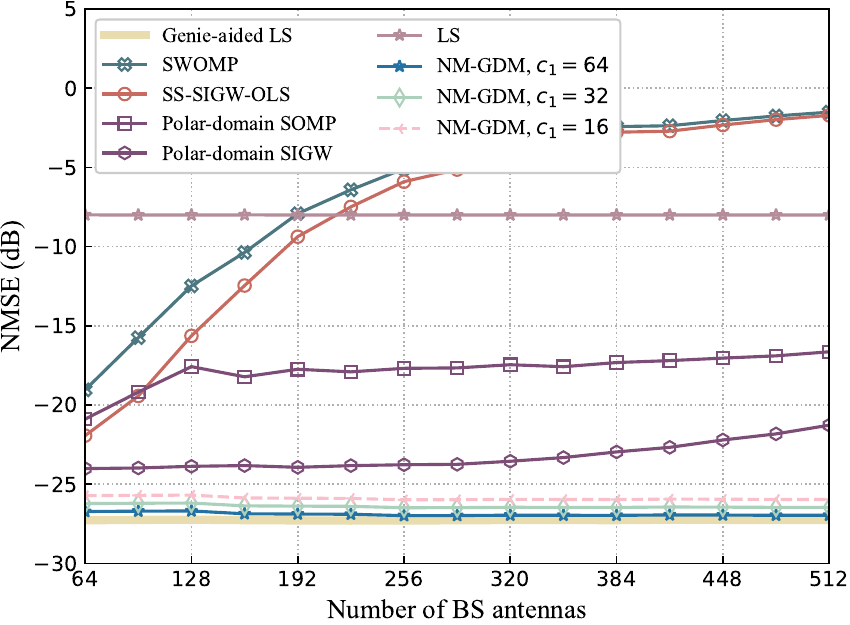}\label{fig:an_r=10}}
	\hfill
	\subfloat[Users are located at (350 m, 400 m).]{\includegraphics[width=0.33\textwidth]{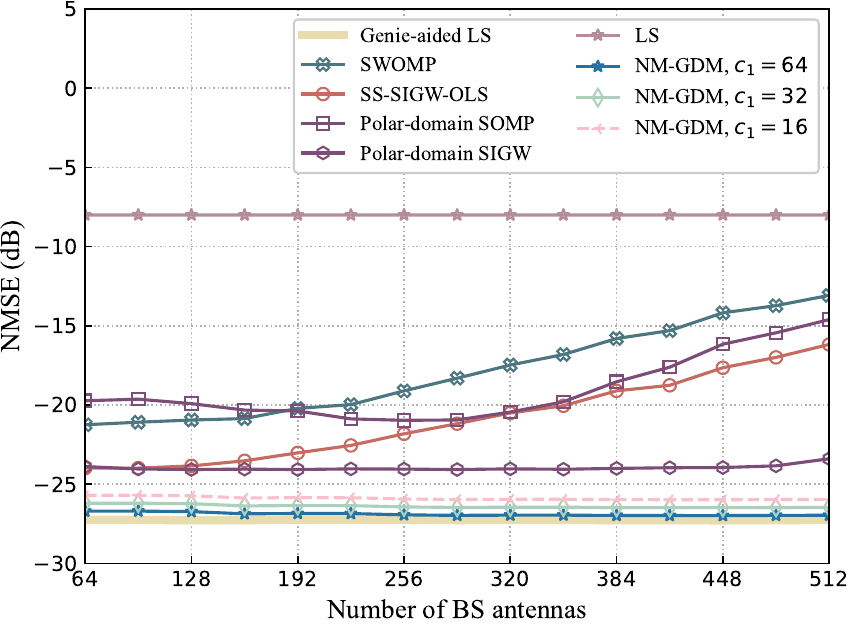}\label{fig:an_r=380}}
	\hfill
	\subfloat[Users are located at (5 m, 35 m).]{\includegraphics[width=0.33\textwidth]{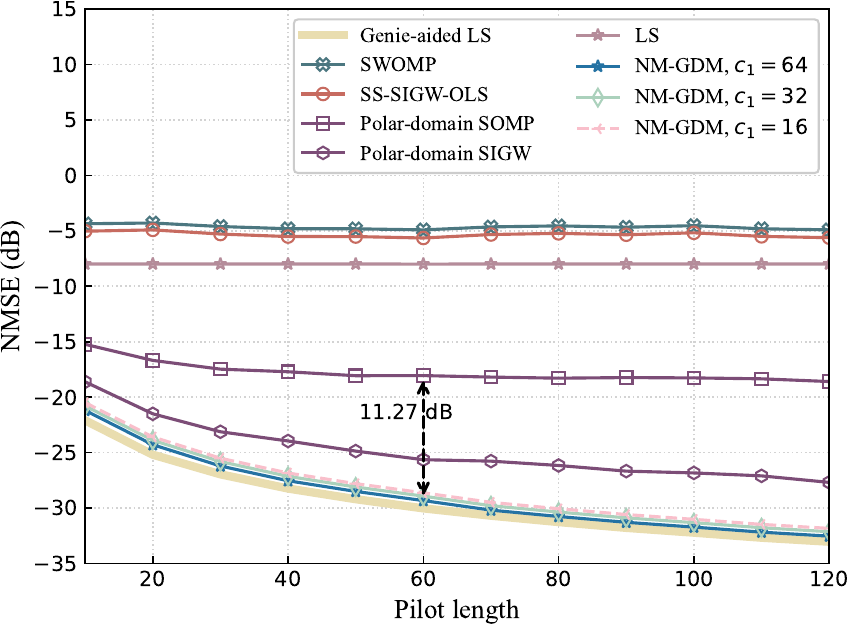}\label{fig:an_pilot}}
	\captionsetup{font=footnotesize}
	\caption{NMSE performance of CE in relation to the number of BS antennas and pilot overhead, respectively.}
	\label{fig:an}
\end{figure*}

\subfigref{fig:snr}{fig:snr_r=18} presents the NMSE of our algorithm and the baselines with respect to SNRs, where the pilot length $Q$ is 32 and the distances between the users and the BS are randomly sampled from the range (15 m, 25 m), i.e., located in the NF region. It can be observed that the NM-GDM method achieves competitive CE accuracy over the entire SNR range. Compared with these baseline methods, the proposed method maintains competitive CE accuracy even at low SNR. This is because the proposed NM-GDM has learned well the implicit prior information (e.g., mean, variance) of the channel distribution during the training so that it can work effectively in the case of low SNR. Besides, the CE performance of the NM-GDM is also robust in the FF region, where users are located within (350 m, 400 m), as shown in \subfigref{fig:snr}{fig:snr_r=398}. As shown in \figref{fig:SNR_PILOT}, we conduct additional experiments to simulate a more practical scenario, where the number of pilot signals is set to 8. The results show that our proposed approach maintains competitive estimation accuracy even under low communication overhead conditions. It can be observed that our approach outperforms the Genie-aided LS algorithm under low-SNR conditions by leveraging learned prior information to effectively mitigate noise. However, in high-SNR regimes, this advantage diminishes, as noise is no longer the dominant factor. Additionally, the proposed NM-GDM, being a learning-based approach, can suffer from inherent generalization errors, stemming from factors such as training strategies and the diversity of the training channel data.




\begin{figure}[!t]
	\centering
	\includegraphics[width=0.38\textwidth]{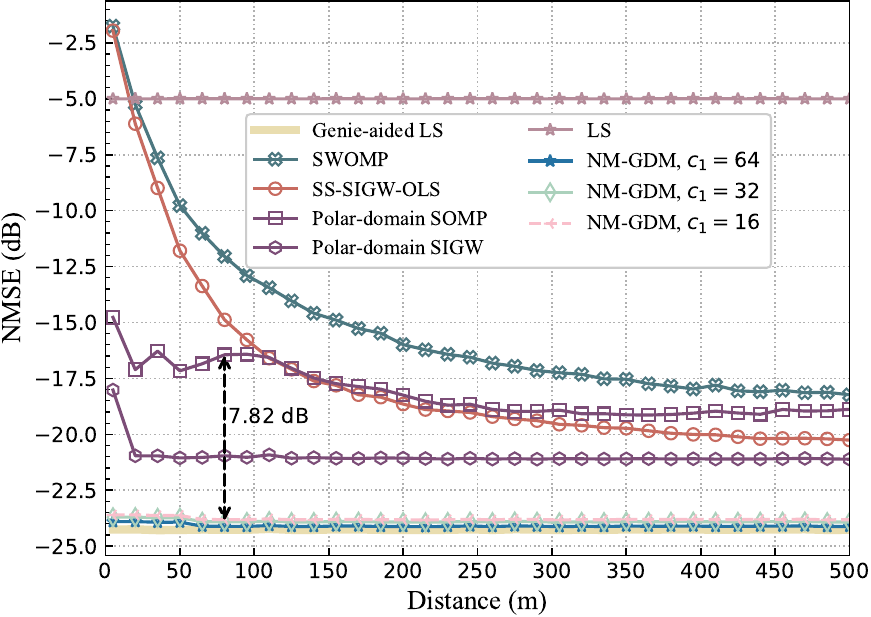}
	\captionsetup{font=footnotesize}
	\caption{Comparison of NMSE performance between the NM-GDM and the baselines at different distances}
	\label{fig:dis}
\end{figure}

To further validate the robustness of the NM-GDM, we vary the number of BS antennas and the distance to strengthen or weaken their corresponding NF effects. \figref{fig:an}(a) and (b) show the NMSE performance of the NM-GDM and the baselines in relation to the count of BS antennas, where the SNR is 8 dB, and the users are located (15 m, 25 m) and (350 m, 400 m), respectively. It is observed that the NMSE performance of the FF algorithms based on the angle domain only is significantly degraded. As shown in \subfigref{fig:an}{fig:an_r=10} and \subfigref{fig:an}{fig:an_r=380}, when the NF effect becomes more pronounced, the performance degradation trend of the baselines become more apparent, especially for the angle domain based-method. Additionally, the proposed approach exhibits minor fluctuations as the number of antennas varies while closely approaching the performance of the desirable Genie-aided LS algorithm. This validates that the proposed algorithm achieves competitive and stable estimation performance across various antenna configurations, demonstrating its robustness, which is attributed to its powerful implicit prior learning capabilities and resilient latent representations.

\subfigref{fig:an}{fig:an_pilot} displays the NMSE performance of the NM-GDM and baselines with respect to the pilot length. The curves in the figure all show a decreasing trend as the pilot length increases, except for the LS algorithm. Across varying pilot lengths, polar-domain algorithms consistently outperform angle-domain methods in NMSE, as the latter doesn't account for NF effects. Our approach surpasses the baseline methods, demonstrating that NM-GDM effectively recovers the channel with reduced communication overhead. \figref{fig:dis} compares the NMSE performance versus distance for the proposed NM-GDM and existing algorithms. Due to the antenna number $N=256$ and frequency $f_c=28$ GHz, the Fraunhofer distance is approximately 351 meters. As the distance decreases, existing algorithms exhibit a deteriorating trend in NMSE performance, except for the LS method, especially when the distance is significantly less than the Fraunhofer distance. In contrast, the NMSE performance of the proposed NM-GDM surpasses the existing algorithms, and remains robust even when the distance is extremely small.
\section{Conclusion}
In this paper, we investigated NF CE in XL-MIMO systems. Considering the increasingly pronounced NF effects, we introduced a joint angle-distance domain spherical-wavefront physical channel model that effectively captures the inherent sparsity of NF channels. By leveraging the sparsity of the NF channel in the joint domain, we reformulated the task as a sparse signal reconstruction problem and introduced the SOMP algorithm to solve it. However, since the SOMP algorithm assumes that the angles and distances align exactly with the sampling grids, this assumption may limit the accuracy of the CE performance.

To mitigate this limitation, we proposed a GenAI-enabled channel refinement approach. While traditional GDM possesses strong implicit prior learning capabilities, its original architecture restricts it to functioning solely as a channel generator, preventing its direct application in CE. Therefore, we developed a side-information-guided GDM for NF channel refinement. By incorporating the initial CE as side information, it gradually guides the GDM's generation process, making this process more controllable. To further reduce the sampling steps in the generation process, we abandoned the Markov chain decomposition scheme and introduced a non-Markovian GDM to accelerate the generation process. Through theoretical analysis, both approaches are shown to share a consistent objective. Experimental results have demonstrated that the proposed approach achieves competitive CE accuracy compared to the baselines, even in scenarios with low SNR or limited communication overhead.


Our work pioneers the use of deep generative diffusion model for XL-MIMO NF CE, and the preliminary results highlight its promising potential for enabling future wireless communications. However, there are still two critical aspects that warrant further exploration. We believe that optimizing the model for a lightweight design and enhancing sampling efficiency are essential factors for enabling the effective deployment of GDM in future wireless communication networks. For model lightweighting, mature techniques such as layer pruning and multi-objective knowledge distillation can be employed to further reduce memory consumption and latency. Moreover, to improve sampling efficiency, consistency constraints or more deterministic sampling processes \cite{song2023consistency, wang2024sinsr} can be introduced, allowing the model to directly predict denoised results at any given time step. Therefore, in future research, we will focus on further exploring and advancing this promising direction.

\bibliographystyle{IEEEtran}
\bibliography{EE_AI}


\end{document}